\documentclass[manuscript=article]{achemso}

\usepackage[version=3]{mhchem} % Formula subscripts using \ce{}
\usepackage{siunitx}
\usepackage{amsmath}
\usepackage{booktabs}
\usepackage{braket}
\usepackage{graphicx}
\usepackage{subcaption}
\usepackage{epstopdf}
\usepackage{amsmath}
\usepackage{color}
\usepackage{upgreek}
\usepackage[version=3]{mhchem} % Formula subscripts using \ce{}
\usepackage[T1]{fontenc}       % Use modern font encodings

\setkeys{acs}{usetitle = true}

\newcommand{\beginsupplement}{%
        \setcounter{table}{0}
        \renewcommand{\thetable}{S\arabic{table}}%
        \setcounter{figure}{0}
        \renewcommand{\thefigure}{S\arabic{figure}}%
     }
\author{Gabriel Sauter}
\altaffiliation{Contributed equally}
\affiliation[Univeristy Heidelberg]
{Physikalisch-Chemisches Institut, Universität Heidelberg, Im Neuenheimer Feld 253,
69120 Heidelberg, Germany}
\author{Antonia Papapostolou}
\affiliation[Univeristy Heidelberg2]
{Interdisziplinäres Zentrum für Wissenschaftliches Rechnen, Universität Heidelberg, Im
Neuenheimer Feld 205A, 69120 Heidelberg, Germany}
\altaffiliation{Contributed equally}
\author{Audrey Pollien}
\affiliation[Univeristy Heidelberg]
{Physikalisch-Chemisches Institut, Universität Heidelberg, Im Neuenheimer Feld 253,
69120 Heidelberg, Germany}
\author{Sergius Boschmann}
\affiliation[Univeristy Heidelberg]
{Physikalisch-Chemisches Institut, Universität Heidelberg, Im Neuenheimer Feld 253,
69120 Heidelberg, Germany}
\author{Kathleen Fuchs}
\affiliation[UniHd3]
{Organisch-Chemisches Institut, Universität Heidelberg, Im Neuenheimer Feld 270, 69120
Heidelberg, Germany}
\author{Jan Freudenberg}
\affiliation[UniHd3]
{Organisch-Chemisches Institut, Universität Heidelberg, Im Neuenheimer Feld 270, 69120
Heidelberg, Germany}
\author{Uwe Bunz}
\affiliation[UniHd3]
{Organisch-Chemisches Institut, Universität Heidelberg, Im Neuenheimer Feld 270, 69120
Heidelberg, Germany}
\author{Andreas Dreuw}
\affiliation[Univeristy Heidelberg2]
{Interdisziplinäres Zentrum für Wissenschaftliches Rechnen, Universität Heidelberg, Im
Neuenheimer Feld 205A, 69120 Heidelberg, Germany}
\author{Petra Tegeder}
\affiliation[Univeristy Heidelberg]
{Physikalisch-Chemisches Institut, Universität Heidelberg, Im Neuenheimer Feld 253,
69120 Heidelberg, Germany}
\email{tegeder@uni-heidelberg.de}
\phone{+49 (0) 6221 548475}

\title[Manuscript]
  {Exceptionally High Two-Photon Absorption in Diazaacene-Bithiophene Derivatives: A Combined Experimental and Theoretical Approach}

\abbreviations{IR,NMR,UV}
\begin{document}

\begin{abstract}
    This study delves into the enhancement of two-photon absorption (2PA) properties in diazaacene-bithiophene derivatives through a synergistic approach combining theoretical analysis and experimental validation. By investigating the structural modifications and their impact on 2PA cross sections, we identify key factors that significantly influence the 2PA efficiency. For all molecular systems studied, our state-of-the-art quantum chemical calculations show a very high involvement of the first excited singlet state (S\textsubscript{1}) in the 2PA processes into higher excited states, even if this state itself has only a small 2PA cross section for symmetry reasons. Consequently, both the oscillator strength of S\textsubscript{1} and the transition dipole moments between S\textsubscript{1} and other excited states are of importance, underscoring the role of electronic polarizability in facilitating effective two-photon interactions. The investigated compounds exhibit large 2PA cross sections over a wide near-infrared spectral range reaching giant values of $\sigma_{2,\text{max}}=42000 \; \text{GM}$. The introduction of diazine and diazaacene moieties into bithiophene derivatives not only induces charge transfer but also opens up pathways for the creation of materials with tailored nonlinear optical responses, suggesting potential applications in nonlinear optics.

\end{abstract}

%%%%%%%%%%%%%%%%%%%%%%%%%%%%%%%%%%%%%%%%%%%%%%%%%%%%%%%%%%%%%%%%%%%%%
%% Start the main part of the manuscript here.
%%%%%%%%%%%%%%%%%%%%%%%%%%%%%%%%%%%%%%%%%%%%%%%%%%%%%%%%%%%%%%%%%%%%%
\section{Introduction}

Two-photon absorption (2PA) represents a cornerstone in the field of nonlinear optics, characterized by the simultaneous absorption of two photons by a molecule to transition into an excited state. This phenomenon stands in contrast to linear one-photon absorption (1PA), offering unique advantages. The nonlinear nature not only enables heightened spatial selectivity, occurring at the focal point with sufficient photon flux, but also ensures tightly confined spatial excitation due to its quadratic dependence on light intensity. Moreover, 2PA's utilization of longer wavelengths for excitation, specifically in the near-infrared (NIR) spectral range, allows for deeper penetration into biological tissues without significant scattering or absorption. Those advantages are particularly beneficial in  high-resolution microscopy and bioimaging\cite{bioimaging, bioimg1,bioimg2,bioimg3}, allowing for detailed visualization of biological processes with minimal background noise and reduced tissue damage. Additionally the precise delivery of energy gives rise to applications in various areas such as nanofabrication processes\cite{nanofab}, targeted photodynamic therapy\cite{phototherapy} and advanced optical data storage solutions\cite{datastorage} that are unattainable through conventional optical techniques.

\begin{figure}[H]
\begin{center}
\includegraphics[width=12cm]{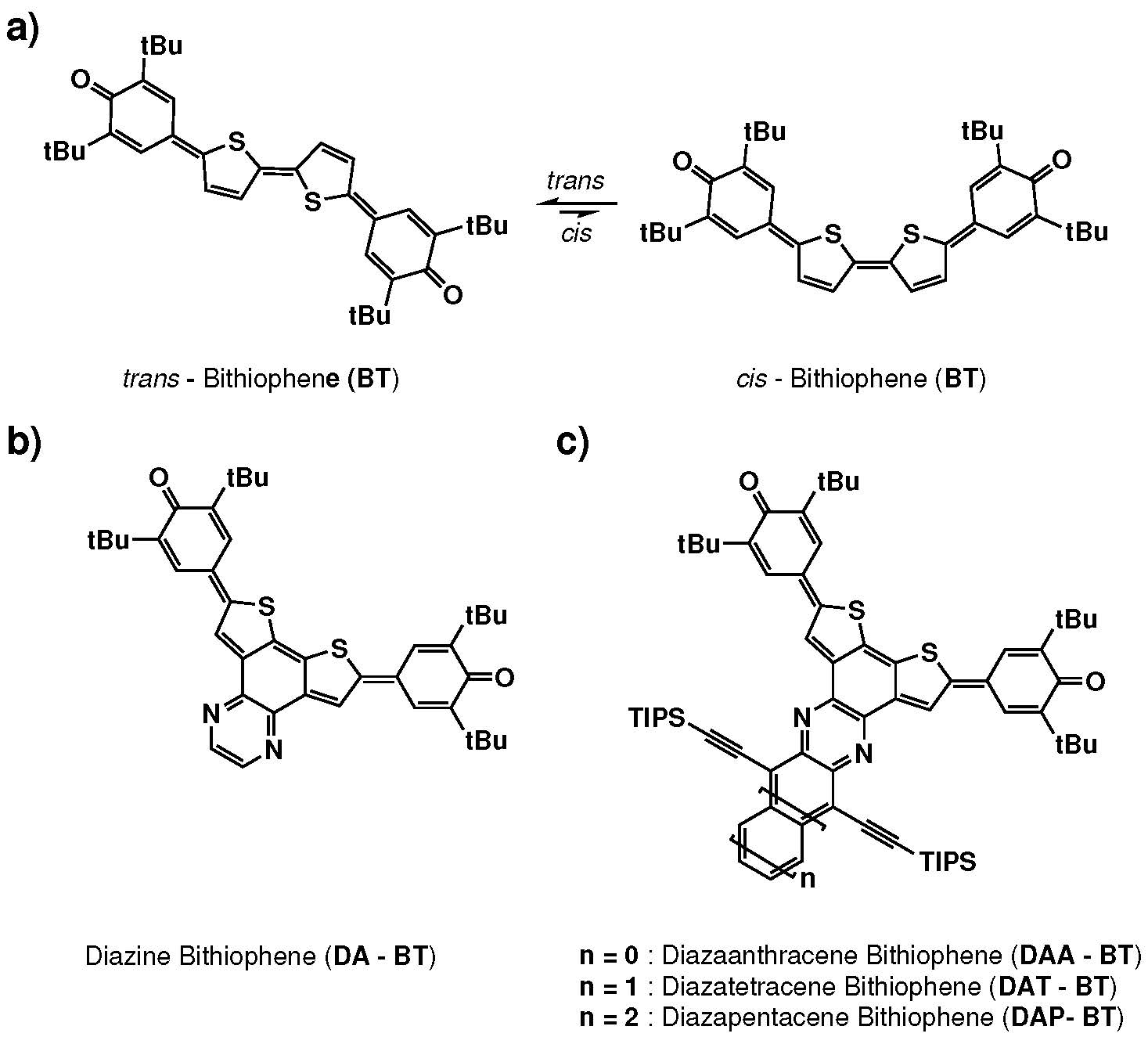}
\caption{Molecules under investigation and their nomenclature: a) Bithiophene (BT) in \textit{trans}- and \textit{cis}-configuration, b) Diazine Bithiophene (DA-BT), and c) Diazaacene Bithiophenes with different acene length from n=0 (DAA-BT) over n=1 (DAT-BT) to n=2 (DAP-BT).}
\label{fig:molecules}
\end{center}
\end{figure}

Aiming at applications in photonics, a great effort has been put into synthesizing materials with high two-photon absorption cross sections ($\sigma_2$) as well as understanding the relationship between the molecular structure and the nonlinear optical processes. In this context, our study focuses on the enhancement of 2PA properties within organic molecules, particularly targeting the diradicaloid diazaacene-bithiophene derivatives\cite{bunzdiradical} (see Fig. \ref{fig:molecules}) that have recently been found to exhibit high nonlinear responses.\cite{Azaacene}
The bithiophene (BT) core is part of a series of heteroquaterphenoquinones, which have demonstrated promising properties in previous studies.\cite{bithio} In particular, the sulfur derivative exhibited an intense absorption band in the near-infrared region, suggesting its potential utility in high-density optical storage materials for diode lasers and other applications.
Another significant aspect of our investigation is the incorporation of N-heteropolycyclic aromatic compounds. This modification not only serves as an efficient electron acceptor but also facilitates the tuning of the electronic structure of the molecules. The presence of nitrogen in the aromatic backbone introduces a push-pull electronic effect, significantly influencing the 2PA cross sections by enhancing the polarizability and electronic responsiveness of the molecule.
Our approach combines both quantum chemical analyses and experimental methodologies to shed light on the factors that govern the 2PA characteristics of these molecules, with the ultimate goal of optimizing their performance for NIR-2PA applications.

Our experimental investigations reveal that the 2PA cross sections of these molecular systems rank among the highest reported for organic compounds, underscoring their potential in applications requiring high 2PA efficiency.\cite{record2pa, overview2pa} Concurrently, our theoretical studies provide a deeper understanding of the 2PA mechanisms in the diazaacene-bithiophene derivatives, employing a sum-over-states (SOS) approach to elucidate the underlying principles. If two photons with the same energy ($\frac{\omega_f}{2}$) are absorbed, the so-called SOS expression for a component of the 2PA transition moment between the ground state and the final state $f$ reads
\begin{equation}
\label{eq:sos}
    S^f_{AB} = \sum_{n\neq 0} \left( \frac{\braket{0|\mu_A|n}\braket{n|\overline{\mu}_B|f}}{\omega_n - \frac{\omega_f}{2}} + \frac{\braket{0|\mu_B|n}\braket{n|\overline{\mu}_A|f}}{\omega_n - \frac{\omega_f}{2}}\right).
\end{equation}
Here, $A,B\in {x,y,z}$ denote the Cartesian components of the electric dipole operator $\vec{\mu}$. As can be seen from Eq.\ (\ref{eq:sos}), several properties contribute to the 2PA cross sections: the transition dipole moments between ground state and excited states ($\braket{0|\vec{\mu}|n}$), the transition dipole moments between final state and other excited states (for $n\neq f$: $\braket{n|\vec{\mu}|f}$) and the difference dipole moment between the ground state and the final state (for $n=f$: $\braket{f|\vec{\overline{\mu}}|f}$) as well as the difference between the excitation energies and the energy of the incident photons ($\omega_n - \frac{\omega_f}{2}$). These considerations will make it possible to investigate the excited states involved in the 2PA processes more closely in the following.

\section{Results and discussion}
The 2PA cross sections were determined by the so-called Z-scan method (see Experimental Section). In addition, they were calculated at the time-dependent density functional theory (TDDFT) level using the SOS approach (see Computational Methodology). The most notable characteristic of all two-photon spectra of the molecular systems (see Fig. \ref{fig:molecules}) in benzene is the presence of two distinct bands at $\approx \SI{1.6}{\electronvolt}$ (“low-energy band”) and at $>\SI{2.6}{\electronvolt}$ (“high-energy band”), as summarized in Table \ref{tbl:bithi}.
The measured Z-scan traces in the higher energy domain exhibit an overlap of a concave curve with a convex one, suggesting the simultaneous occurrence of 2PA and saturable absorption (SA) - as further elaborated in the Supplementary Information. This overlap complicates the identification and determination of the exact maximum position of the 2PA band, given the nonlinear competing effect of SA. Consequently, in Table \ref{tbl:bithi}, the energy values for the high-energy transition carry a minimum estimation uncertainty.

\begin{table}[]
  \caption{Absorption maximum of the one-photon excitation for the investigated systems in benzene. For all compounds, the maximum linear absorption lays in the low-energy band.}
  \label{tbl:1pa}
  \begin{tabular}{@{}lrrlrr@{}}
    \toprule
    & \multicolumn{2}{c}{Low-energy band} && \multicolumn{2}{c}{High-energy band} \\

     \cmidrule(lr){2-3} \cmidrule(lr){5-6}
    & $E_{\text{a}}$\textsuperscript{\emph{a}} & $\lambda_{\text{a}}^{(1)}$ & &$E_{\text{b}}$\textsuperscript{\emph{a}} &  $\lambda_{\text{b}}^{(1)}$ \\
    \midrule

    BT&  \SI{1.83}{\electronvolt} & \SI{678}{\nano\meter} && & \\
    DA-BT&  \SI{1.75}{\electronvolt} & \SI{708}{\nano\meter} && \SI{2.57}{\electronvolt} & \SI{482}{\nano\meter} \\
    %DA(n)-BT &  &  &  &  \\
    DAA-BT & \SI{1.72}{\electronvolt} & \SI{721}{\nano\meter} && \SI{3.12}{\electronvolt} & \SI{7200}{\nano\meter} \\

    DAT-BT & \SI{1.71}{\electronvolt} & \SI{725}{\nano\meter} && \SI{2.89}{\electronvolt} & \SI{14900}{\nano\meter} \\
    DAP-BT  & \SI{1.72}{\electronvolt} & \SI{725}{\nano\meter} && \SI{2.57}{\electronvolt} & \SI{19900}{\nano\meter} \\
    \bottomrule
  \end{tabular}

  \textsuperscript{\emph{a}}$E_{\text{a}}$ is the energy of the peak in the absorption band.

\end{table}

\begin{table}[H]
  \caption{Measured 2PA parameters for the investigated systems in benzene.}
  \label{tbl:bithi}
  \begin{tabular}{@{}lrrlrr@{}}
    \toprule
    & \multicolumn{2}{c}{Low-energy band} && \multicolumn{2}{c}{High-energy band} \\

     \cmidrule(lr){2-3} \cmidrule(lr){5-6}
    & $E_{\text{max}}$\textsuperscript{\emph{a}} & $\sigma_{2,\text{max}}$ & &$E_{\text{max}}$\textsuperscript{\emph{a}} &  $\sigma_{2,\text{max}}$ \\
    \midrule

    BT&  \SI{0.86}{\electronvolt} & \SI{5500}{GM} && \SI{1.46}{\electronvolt} & \SI{42000}{GM} \\
    DA-BT&  \SI{0.93}{\electronvolt} & \SI{900}{GM} && \SI{1.32}{\electronvolt} & \SI{6800}{GM} \\
    %DA(n)-BT &  &  &  &  \\
    DAA-BT & \SI{0.77}{\electronvolt} & \SI{1300}{GM} && \SI{1.31}{\electronvolt} & \SI{7200}{GM} \\
    && &&\SI{1.43}{\electronvolt} & \SI{14800}{GM}\\
    DAT-BT & \SI{0.85}{\electronvolt} & \SI{1650}{GM} && \SI{1.33}{\electronvolt} & \SI{14900}{GM} \\
    DAP-BT  & \SI{0.78}{\electronvolt} & \SI{6900}{GM} && \SI{1.30}{\electronvolt} & \SI{19900}{GM} \\
    \bottomrule
  \end{tabular}

  \textsuperscript{\emph{a}}$E_{\text{max}} = \frac{hc}{\lambda_{\text{in}}}$ is the energy of the incident photons.

\end{table}

\begin{figure}
\begin{center}
\includegraphics[width=14cm]{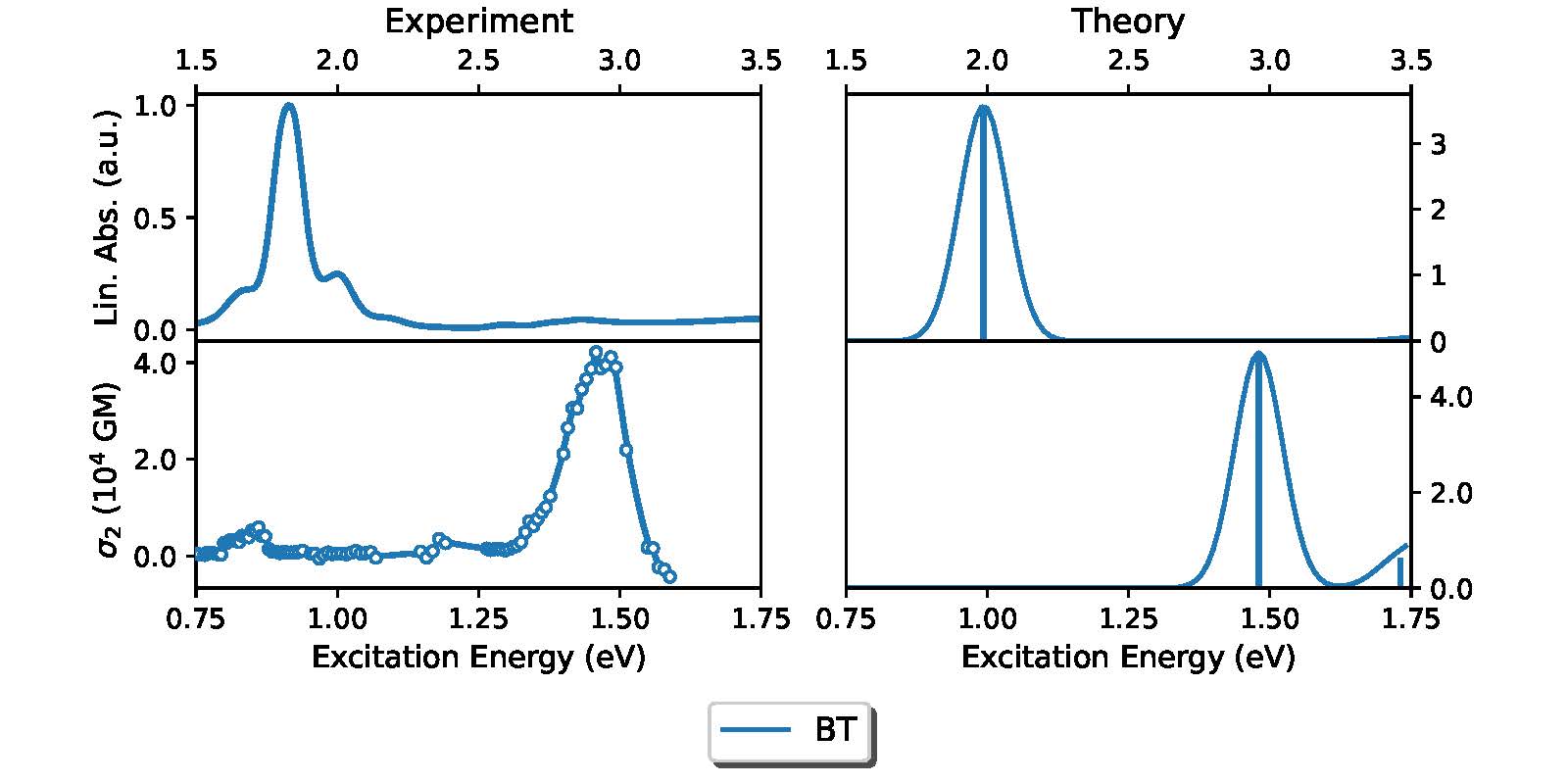}
\caption{Experimental and theoretical  analysis of the 1PA and 2PA characteristics of BT in benzene.}
\label{fig:BT}
\end{center}
\end{figure}

\begin{figure}[H]
    \centering
    %\begin{subfigure}{\textwidth}
        %\centering
        \includegraphics[width=0.8\linewidth]{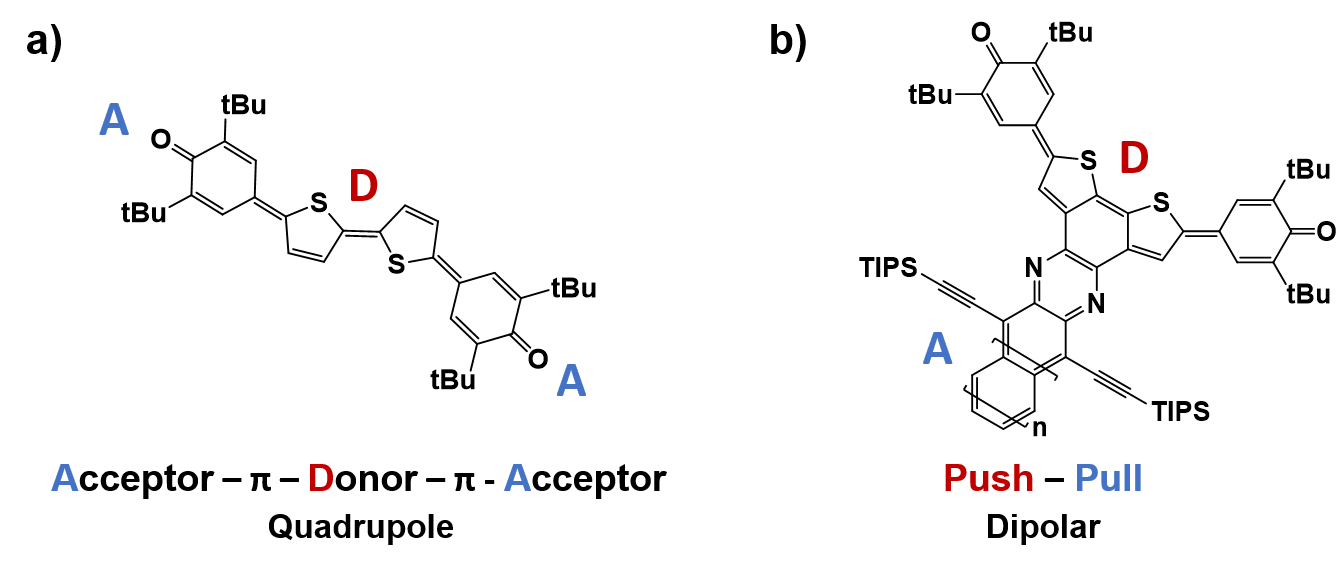}
        %\caption{}
        %\label{fig:QuadCT1}
    %\end{subfigure}

    %\begin{subfigure}{\textwidth}
    %    \centering
    %    \includegraphics[width=0.35\linewidth]{Figs/ad_BT.png} % Replace antonia
    %    \caption{TODO-Antonia States Quadrupole BT and d) CT from Acene-Sulfur}
    %    \label{fig:BT2}
    %\end{subfigure}

    \caption{The BT geometry offers a quadrupolar A-$\pi$-D-$\pi$-A configuration in a). b) The additional azaacene moiety gives rise to additional electron withdrawing.}
    \label{fig:QuadCT-all}
\end{figure}

\paragraph{2PA Properties of BT}
BT is the core component of the DA-BT derivatives\cite{bunzdiradical}. As shown by Takahashi et al., BT (heteroquaterphenoquinone) exists predominantly as the \textit{trans}-isomer in solution.\cite{bithio} This will result in a centrosymmetric system. Notably, it is well-established that the selection rules governing one-photon absorption and two-photon absorption are mutually exclusive for centrosymmetric molecules such as BT. Thus, for symmetry reasons, a state that is bright in the 1PA spectrum must be dark in the 2PA spectrum.
In the experimental UV-vis data depicted in Figure \ref{fig:BT}, a distinct 1PA band is observed at \SI{1.66}{\electronvolt} (\SI{1.82}{\electronvolt} and \SI{2.01}{\electronvolt}). This peak corresponds to an excitation into the first excited singlet state and is faithfully reproduced in the calculated spectrum, albeit with a slight blueshift of about \SI{0.2}{\electronvolt}. The shoulders can be attributed to vibronic contributions and are therefore not present in the calculated spectrum, since only electronic contributions were taken into account.
Despite the inversion symmetry, this first excitation is not completely dark in the experimental 2PA spectrum, which is due to several factors: the presence of both \textit{trans}- and \textit{cis}-configurations of BT in the measured solution (see Figure S1 (SI)) as well as vibrational distortions that relax the selection rules.
When examining the higher energy regime, the situation is exactly the opposite: there is no band in the linear absorption spectrum, but an extremely high peak in the 2PA spectrum, which can be assigned to the second excited state. In the experiment, this transition appears at a photon energy of \SI{1.45}{\electronvolt}, consistent with the calculated spectrum at \SI{1.48}{\electronvolt}. This high-energy peak reaches $\sigma_2 = 42000$ GM, which documents a giant 2PA cross section compared to organic molecules of similar size\cite{record2pa,overview2pa}. The BT unit has a large acceptor–$\pi$–donor–$\pi$–acceptor (A–$\pi$–D–$\pi$–A)-type structure as illustrated in Figure \ref{fig:QuadCT-all} a. The resulting quadrupolar structure typically involves a central donor (D) moiety flanked by $\pi$-conjugated linkers and terminal acceptor (A) groups. It has been demonstrated that conjugated molecules, when substituted with donor or acceptor groups in an essentially centrosymmetric pattern, exhibit significantly enhanced two-photon cross sections, often surpassing the corresponding unsubstituted molecules by at least an order of magnitude.\cite{Wang_2001,ICT,ICTPI-Bi,ICT-Bi}

\begin{figure}[H]
\begin{center}
\includegraphics[width=0.5\textwidth]{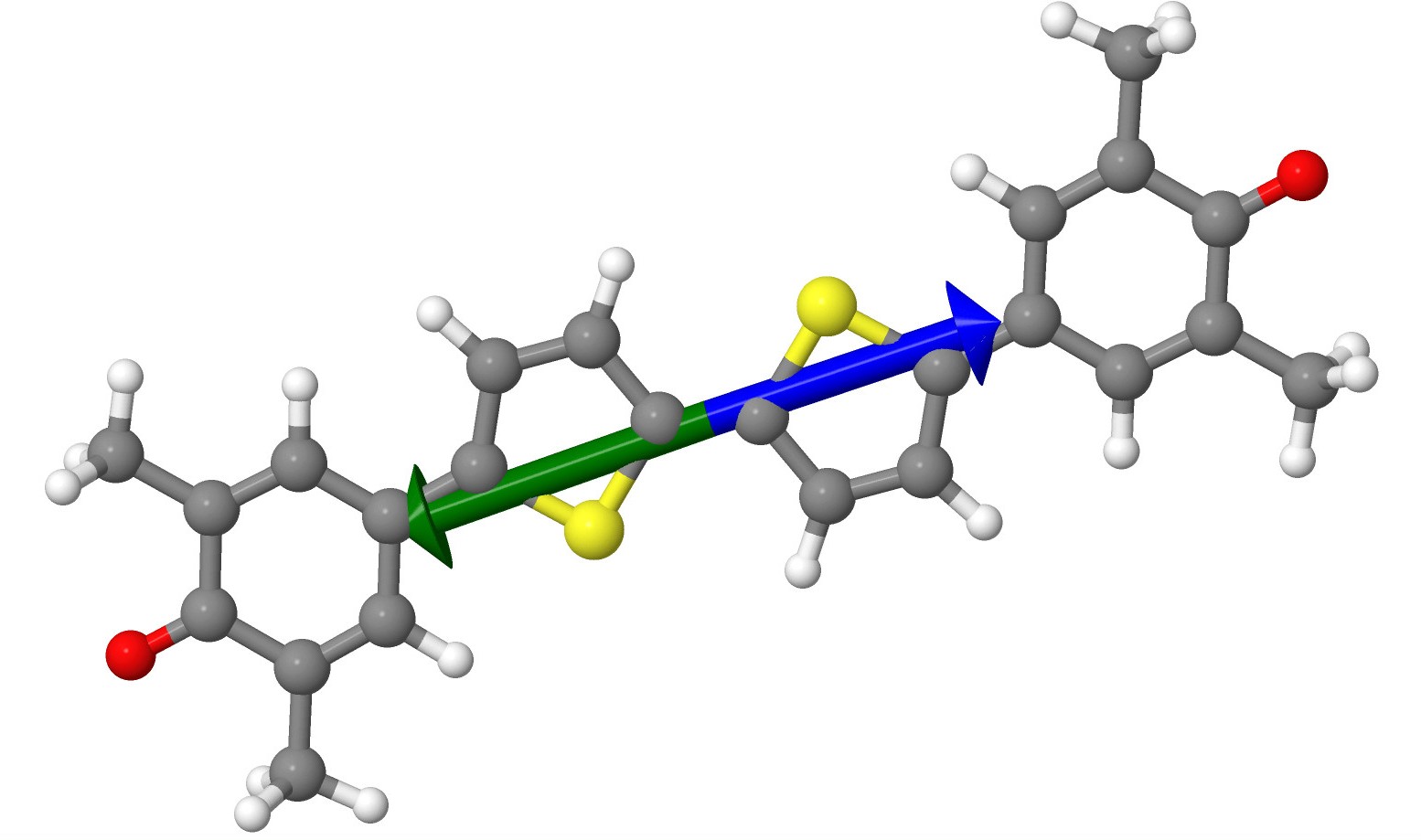}
\caption{Visualization of $\braket{0|\vec{\mu}|1}$ (green) and $\braket{1|\vec{\mu}|2}$ (blue) transition dipole in BT.}
\label{fig:BT_01_12}
\end{center}
\end{figure}

\begin{figure}[H]
\begin{center}
\includegraphics[width=0.4\textwidth]{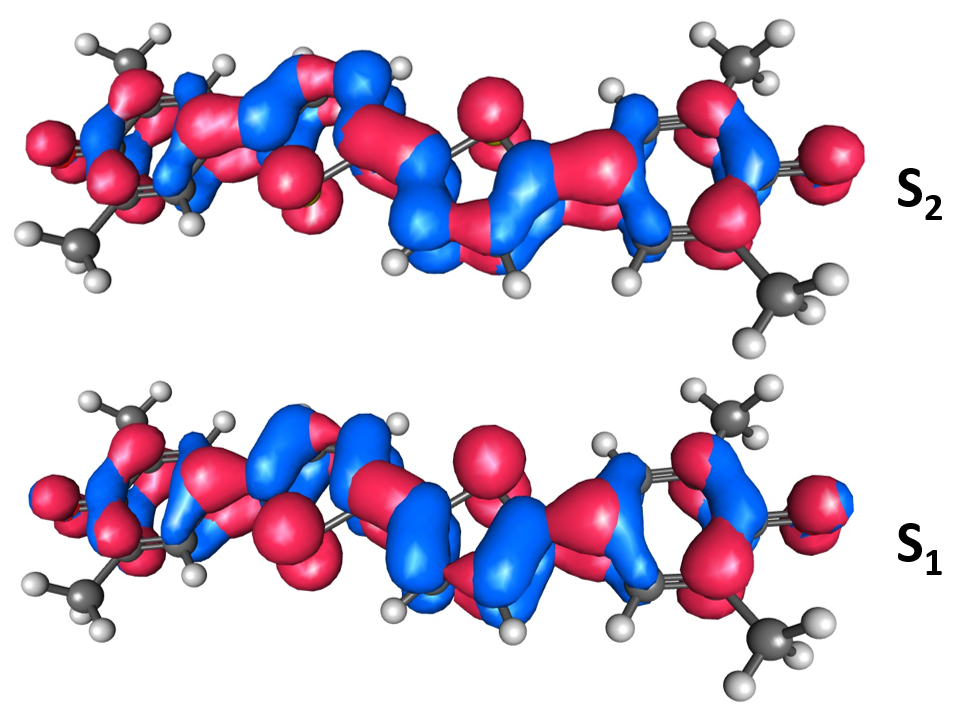}
\caption{Attachment/detachment densities in blue/red of the first two excited singlet states of BT calculated at the CAM-B3LYP-D3(BJ)/aug-cc-pVDZ level of theory. The other excited states can be found in the Supplementary Information.}
\label{fig:ad_BT}
\end{center}
\end{figure}

Figure \ref{fig:BT} shows that the calculations at the CAM-B3LYP-D3(BJ)/aug-cc-pVDZ level of theory not only reproduce the shape of the experimental 1PA spectrum very well, but also that of the experimental 2PA spectrum. According to Beerepoot et al.,\cite{Beerepoot_2018} range-separated functionals such as CAM-B3LYP are well suited to correctly predict qualitative trends of 2PA strengths, even if the underestimation of excited-state dipole moments as well as the overestimation of excitation energies leads to a tendency for the absolute values to be too low, albeit in a rather systematic way.
In order to gain a deeper insight into the 2PA processes of the BT molecule and thus explain the extraordinarily high 2PA cross section of the second excited state, the individual contributions of the SOS expression must be investigated. For this purpose, the calculated molecular properties relevant for this study are summarized in Table \ref{tbl:sos_bt}. Resonance effects can still be ruled out in the energy range under consideration, which is why the excitation energy $\omega_f$ is of rather minor importance in this analysis. Since the states with the large $\sigma_2$ themselves have an oscillator strength close to zero, the terms for $n=f$ in the sum in Eq.\ (\ref{eq:sos}) cannot make a large contribution, and therefore the difference dipole moments of the excited states do not play a major role either. However, the particularly bright first excited state of the BT molecule in the UV-vis spectrum in Figure \ref{fig:BT} is striking. Consequently, those excited states that have a large transition dipole moment with the first excited state should also have a large 2PA transition moment, since the term for $n=1$ then has a particularly high contribution in the sum in Eq.\ (\ref{eq:sos}), which is confirmed by Table \ref{tbl:sos_bt}. This is especially true if $\braket{0|\vec{\mu}|1}$ and $\braket{1|\vec{\mu}|f}$ are aligned,\cite{Cronstrand_2002} which is the case here for the second excited state (see Figure \ref{fig:BT_01_12}). The high $\sigma_2$ of S\textsubscript{2} thus results from the strongly allowed one-photon transition into S\textsubscript{1} in combination with the strong coupling of the first two excited states with each other. Looking at the attachment/detachment densities of the first and second excited states in Figure \ref{fig:ad_BT}, their great similarity is also noticeable.

\begin{table}[H]
\caption{Excitation energies ($\omega_f$), oscillator strengths ($f_L$), 2PA cross sections ($\sigma_2$), transition dipole moments from the ground state, difference dipole moments, and state-to-state transition moments for the first 5 excited singlet states of BT calculated at the CAM-B3LYP-D3(BJ)/aug-cc-pVDZ level of theory.}
\label{tbl:sos_bt}
\begin{tabular}{ccccccc}
\toprule
%\multicolumn{7}{c}{\textbf{BT}}\\
%\midrule
$f$ & $\omega_f$ [eV] & $f_L$ [a.u.] & $\sigma_2$ [GM] & $\vert\braket{0|\vec{\mu}|f}\rvert$ [a.u.] & $\lvert\braket{f|\vec{\overline{\mu}}|f}\rvert$ [a.u.] & $\lvert\braket{1|\vec{\mu}|f}\rvert$ [a.u.]\\
\midrule
1 & 1.99 & 3.568 & 0 & 8.560 & 0.000 & - \\
2 & 2.96 & 0.000 & 49211 & 0.000 & 0.000 & 7.839 \\
3 & 3.10 & 0.000 & 0 & 0.003 & 0.093 & 0.007 \\
4 & 3.10 & 0.000 & 0 & 0.007 & 0.093 & 0.011 \\
5 & 3.46 & 0.000 & 6163 & 0.001 & 0.005 & 1.240 \\
\bottomrule
\end{tabular}
\end{table}

\begin{figure}[H]
\begin{center}
\includegraphics[width=14cm]{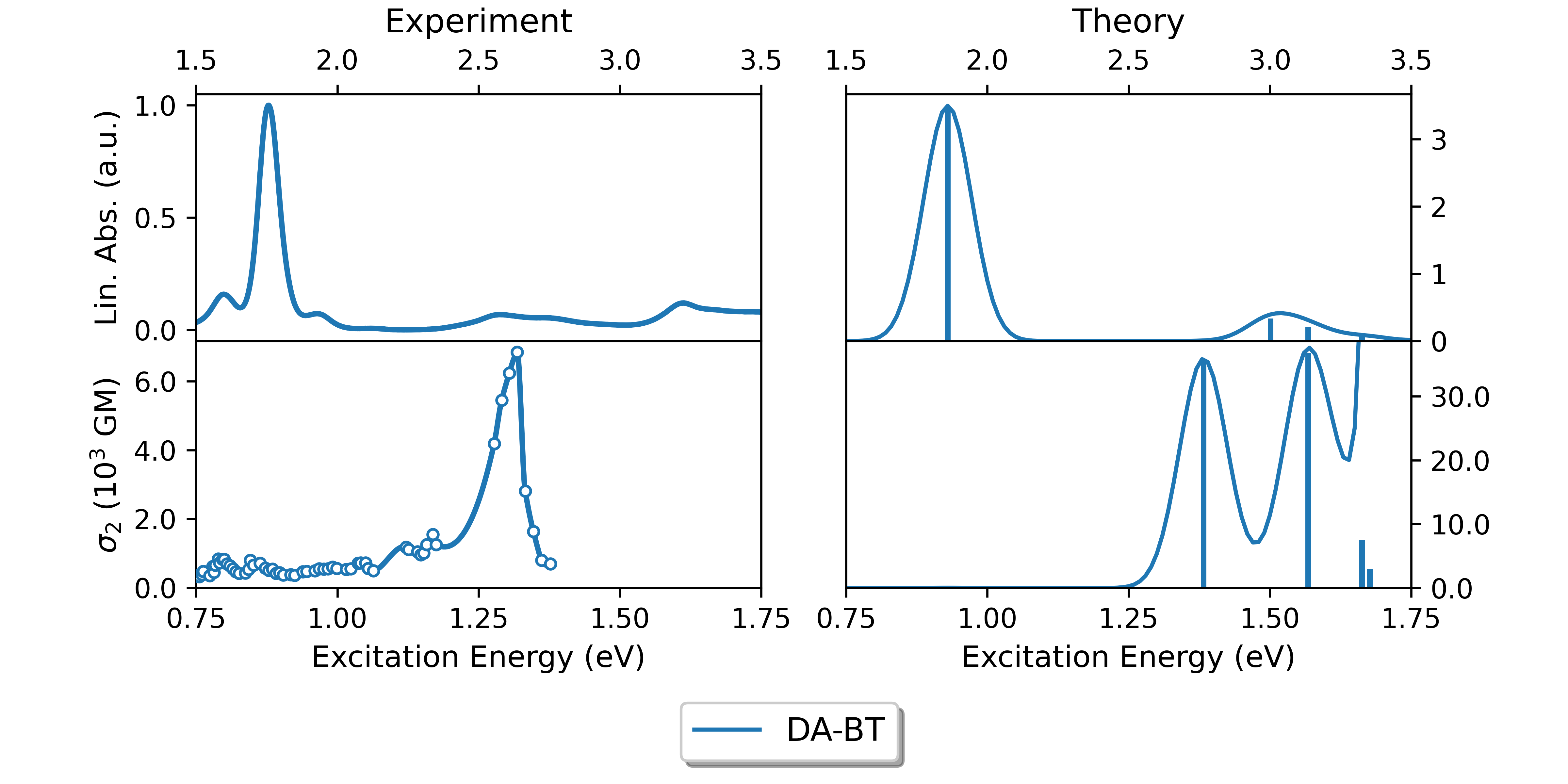}
\caption{Measured and calculated 1PA and 2PA characteristics of DA-BT}
\label{fig:DAfBT}
\end{center}
\end{figure}

\paragraph{Modification with Diazine Moiety (DA-BT)}
Derivatization of the bithiophene with a diazine (DA) moiety, increases the planarity and eliminates the \textit{trans}-configuration, leads to a C\textsubscript{2v} symmetry. This modification introduces an additional dipolar push-pull unit from the electron-donating sulfur site to the newly added electron-withdrawing diazin (see Fig. \ref{fig:QuadCT-all}b). Both the 1PA and 2PA spectra (see Figure \ref{fig:DAfBT}) display similar behavior compared to BT, the peaks are only slightly red-shifted.
Although the molecule has lost the center of symmetry, the first excited state, which still exhibits the largest one-photon transition strength in the UV-vis spectrum at 1.75 eV, has only a small 2PA cross section; the calculated $\sigma_2$ is almost negligible. In contrast, the two-photon transition into the second excited state, which occurs in the experimental spectrum at \SI{1.32}{\electronvolt} and in the theoretical data at \SI{1.39}{\electronvolt}, is again significantly stronger. However, the calculated 2PA peak is much more pronounced than the measured one, indicating experimental limitations in extracting the full $\sigma_2$ due to the competing  linear absorption, which already begins in this range as a result of vibrational broadening.
However, also for DA-BT, the correlation between a high 2PA cross section and a large transition dipole moment between the first and final excited states is evident (see Supporting Information).

\begin{figure}
\begin{center}
\includegraphics[width=14cm]{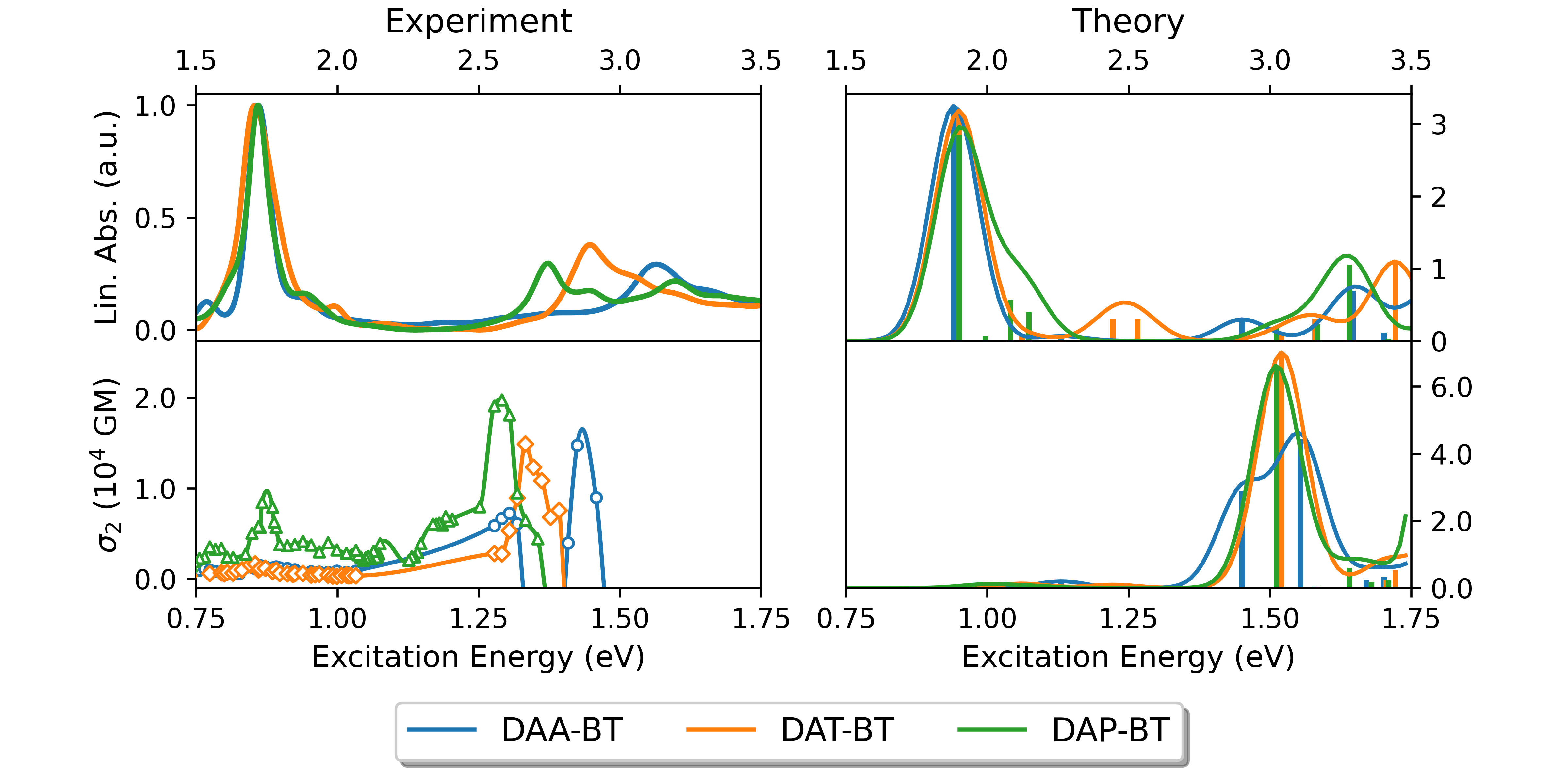}
\caption{1PA and 2PA measurements and calculations of DA(n)-BT with increasing length (n=0-2).}
\label{fig:DAnfBT}
\end{center}
\end{figure}

\begin{figure}
\begin{center}
\includegraphics[width=\textwidth]{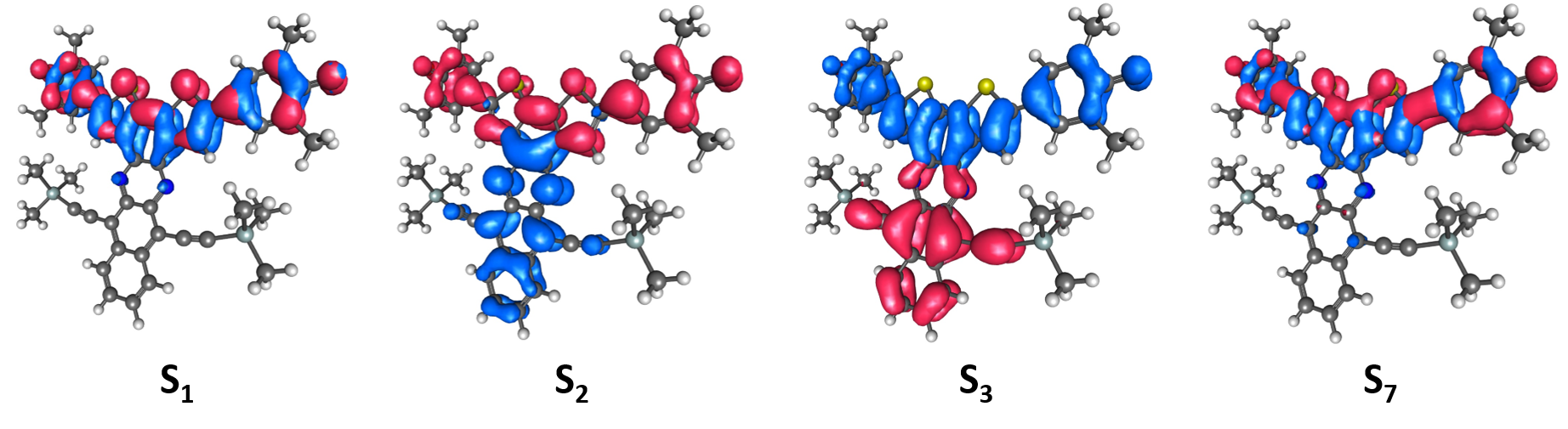}
\caption{Attachment/detachment densities in blue/red of selected excited states of DAT-BT calculated at the CAM-B3LYP-D3(BJ)/aug-cc-pVDZ level of theory.}
\label{fig:ad_DATfBT}
\end{center}
\end{figure}

\paragraph{Extension to Diazaacene Bithiophenes (DA(n)-BT)}
When the diazine is extended by further annulated benzene rings, a strong 2PA can still be observed despite the competing saturable absorption, which occurs at excitation energies of $E_\text{exc} = \SI{1.30}{\electronvolt}$ and above (see Figure \ref{fig:DAnfBT}). For DAA-BT, there is a double peak structure with $\sigma_2(E_{exc}=\SI{1.31}{\electronvolt})=7250 \;\text{GM}$ separated by SA from a second maximum with $\sigma_2(E_{exc}=\SI{1.43}{\electronvolt})=14800\; \text{GM}$.
Further modification by extending the acene length in diazaacene-bithiophene (DA(n)-BT) results in tuning of the 2PA band position. Increasing the acene length from n=0 to n=2 leads to an uneffected magnitude of $\sigma_2$ in the higher energy region and an increase in two-photon absorption in the lower energy region, highlighting the versatility of structural modifications in optimizing photophysical properties.
For the detailed analysis the data of the DAT-BT will be discussed, the corresponding data for the two other compounds can be found in the Supporting Information. If one compares the spectra in Figure \ref{fig:DAnfBT} with the spectra in Figure \ref{fig:BT}, it is noticeable that the two ``low-energy'' and ``high-energy'' bands remain almost unchanged except for a red-shift, since the respective excitations are still localized at the BT unit and the acene has only a small influence, as can be seen from the attachment/detachment densities of S\textsubscript{1} and S\textsubscript{7} (see Figure \ref{fig:ad_DATfBT}).
Compared to BT, however, there are additional low-lying excitations that also involve the acene moiety. The TDDFT calculations show that two new states emerge that are visible in both the 1PA and 2PA spectra. In both cases, these involve charge-transfer excitations that extend over the entire molecule: for S\textsubscript{2} from the BT unit to the acene unit and for S\textsubscript{3} vice versa (see Figure \ref{fig:ad_DATfBT}). Based on its position, one could assume at first glance that the enhanced absorption band between \SI{0.75}{\electronvolt} and \SI{1.00}{\electronvolt} in the experimental 2PA spectra (Figure \ref{fig:DAnfBT}) can be attributed to excitations into the first excited state.
However, since these states are only about 0.1-\SI{0.3}{\electronvolt} higher in energy than S\textsubscript{1} state (depending on the molecule, see Supporting Information), and according to the calculations have a significantly higher 2PA cross section, these are assigned to charge-transfer excitations. A slight shift in peak position between 1PA and 2PA spectra is common due to vibronic effects not accounted for in our calculations.\cite{Lin_2011} In the case of DAP-BT, a $\sigma_2$ of up to 6900~GM can be measured as a result of this charge-transfer excitation. The calculated values are smaller, but a severe underestimation of absolute 2PA transition strengths by range-separated functionals is already known.\cite{Beerepoot_2015, Beerepoot_2018} It can also be observed that the 2PA cross section still correlates with the transition strength between the first excited state and the final state (see Supporting Information). For all states with significant $\sigma_2$ (S\textsubscript{2}, S\textsubscript{3}, and S\textsubscript{7}), $\braket{0|\vec{\mu}|1}$ and $\braket{1|\vec{\mu}|f}$ are aligned. However, since the molecule has no inversion center, the transition dipole moments between the ground state and the final states are no longer negligible, so that the terms for $n=f$ and thus the excited-state dipole moments also have a contribution.

\subsection{Conclusion}

Our comprehensive study on the 2PA properties of bithiophene derivatives emphasizes the significant impact of structural modifications on their photophysical behavior. We have shown that extended A–$\pi$–D–$\pi$–A compounds show large 2PA cross sections peaking at 42000 GM for BT. All molecules under investigation show a significant 2PA (over 100 GM) in a wide range of the low-energy band ($0.75-0.95$ eV). By introducing diazine and diazaanthracene moieties to the BT, we have demonstrated the ability to induce charge transfer, enhance 2PA properties, and tune the absorption band positions in the high-energy band (1.25--1.4 eV), offering promising avenues for the development of advanced materials for NIR-2PA applications. We thus have demonstrated that relatively small changes in molecular structure can lead to a large impact on the optical properties in this class of materials. The observed correlation not only enhances our understanding of the underlying mechanisms governing 2PA but also provides a valuable framework for the design and development of materials with optimized nonlinear optical properties.

\section{Experimental Methods}

\subsection{Two-Photon Absorption Measurements}

\subsubsection{Sample Preparation}
The two-photon absorption (2PA) measurements were conducted on solutions containing the derivatives of interest. These derivatives were dissolved in benzene to reach a concentration of 1.7 mM (for BT), 2 mM (for DA-f-BT) and 1 mM (for DA(n)-f-BT), all under a nitrogen atmosphere to avoid reduction reactions that might alter sample’s properties. The prepared solutions were then placed into 2 mm path length Infrasil cuvettes (Starna 1/ST/C/I/2) to ensure accurate and consistent path length for the measurements.
\subsubsection{Open-Aperture Z-Scan Technique}
The 2PA properties of the solutions were assessed using the open-aperture Z-scan technique, a reliable method for determining nonlinear absorption characteristics.\cite{Sutherland.1996} A picosecond Nd:YAG laser (Ekspla PL2230) operating at a 50 Hz repetition rate was utilized, ensuring complete thermal relaxation between pulses. To cover a broad spectral range in the near-infrared region, the laser beam was modified using a harmonic generator and a parametric generator, achieving a narrow pulse width ($\lambda_p=3.7$ nm). The laser pulses, characterized by a Gaussian temporal profile with a full width at half maximum (FWHM) of 18 ps, were spatially reshaped using a Gaussian filter. The transmittance $T$ for each pulse and z-stage position was recorded by comparing the pulse energy after and before propagation through the sample, using a pyroelectric energy meter (Ophir Optronics PE9-C) for pulse energy measurements.

\subsubsection{Analysis}
The analysis of the open-aperture Z-scan traces focused on the absence of beam distortion, indicating that the transmittance changes were solely due to nonlinear absorption. For a temporally Gaussian pulse, the normalized transmitted energy as a function of sample position $z$ is given by\cite{SheikBahae.1990}:

\begin{equation}
T(z)=  \frac{1}{\sqrt{\pi q_0 (z,0)}}  \int_{-\infty}^{\infty} \ln\left[1+q_0 (z,0)  e^{-t^2 }\right]  dt, \label{eqn:Tz}
\end{equation}

where $q_0 (z,t)=\beta I_0 L_{\text{eff}}/(1+x^2)$, and $x=z/z_0$ with $z_0$ being the Rayleigh length, $\beta$ the nonlinear absorption coefficient, $I_0$ the maximum irradiance, and $L_{\text{eff}}$ the effective sample length. For $|q_0 |<1$, this equation can be simplified to a summation suitable for numerical evaluation:

\begin{equation}
T(z)= \sum_{m=0}^{\infty} \left[\frac{-q_0 (z,0)}{(m+1)^{3/2}}\right].\label{eqn:numerical}
\end{equation}

Curve fitting of the Z-scan traces was performed using this equation. The 2PA cross section $\sigma_2$ was then calculated using the formula:

\begin{equation}
\sigma_2=\frac{h\nu\beta}{N_A d}, \label{eqn:s2}
\end{equation}

where $N_A$ is Avogadro’s constant, $h\nu$ the photon energy, $\beta$ the two-photon absorption coefficient, and $d$ the concentration. The 2PA cross section is reported in Göppert–Mayer (GM) unit (1 GM = $10^{-50}$ cm$^4$ s molecule$^{-1}$ photon$^{-1}$) and was evaluated for every measured excitation wavelength. Measurements at the same wavelength but different input peak irradiances have been performed to exclude multiphoton effects and to verify the nonlinear correction through the fitted beta.

\section{Computational Methodology}
\subsection{Computational Details}
% citations still missing!
All quantum chemical calculations were carried out using a development version of the Q-Chem 6.0 program package.\cite{Epifanovsky_2021} For geometry optimizations, frequency and excited-state calculations, the conductor-like polarizable continuum model (C-PCM) was used to simulate the solvation effects of benzene ($\epsilon=2.3$).\cite{Liu_2013} The geometries of the investigated systems were optimized at the B3LYP-D3(BJ)/6-31G* level of theory. It should be noted that \textit{t}-Bu and TIPS groups were replaced by Me and TMS groups, respectively. Frequency calculations at the same level of theory confirmed in each case that the stationary point is a local minimum. Using the range-separated hybrid exchange-correlation functional CAM-B3LYP\cite{Yanai_2004} with Grimme's D3(BJ) dispersion correction\cite{Grimme_2010,Grimme_2011} and aug-cc-pVDZ as the basis set,\cite{Woon_1993} the 100 lowest excited singlet states were computed within the Tamm-Dancoff approximation (TDA),\cite{Hirata_1999} including the transition moments between excited states and excited-state dipole moments. In order to analyze the electronic structure of the most important states involved in the 2PA processes, attachment and detachment densities were also calculated.

\subsection{Analysis}
The UV-vis spectra were generated by convolving the computed stick spectra with Gaussian functions. The 2PA spectra were obtained from the calculated properties by applying a few-state model. For this purpose, the sum in Eq.\ (\ref{eq:sos}) was formed over the first 100 excited singlet states. Since the denominator in Eq.\ (\ref{eq:sos}) becomes larger with increasing energetic distance between the final state and the other excited states and the corresponding contributions thus become progressively smaller, this approximation is justified. For linearly polarized light with parallel polarization, the rotationally averaged 2PA strength (in atomic units) can be calculated from the obtained tensor as
\begin{equation}
    \langle\delta^{2PA}\rangle = \frac{1}{15}\sum_{AB}(2S_{AB}S_{AB} + S_{AA}S_{BB}).
\end{equation}
According to Ref.\ \citenum{Beerepoot_2015}, the macroscopic 2PA cross section (in cgs units), which is compared with the experiment, can then be determined as follows
\begin{equation}
    \sigma_2=\frac{4\pi^3\alpha a_0^5\omega^2}{c}\langle\delta^{TPA}\rangle g(2\omega,\omega_f,\Gamma),
\end{equation}
where $\alpha$ is the fine structure constant, $a_0$ the Bohr radius, $\omega$ the photon energy in atomic units, $c$ the speed of light, and $g(2\omega,\omega_f,\Gamma)$ the lineshape function that describes spectral broadening effects. In our case, a Gaussian lineshape function
\begin{equation}
    G(2\omega)=\frac{\sqrt{ln2}}{\Gamma\sqrt{\pi}}exp\left[-ln2\left(\frac{2\omega-\omega_f}{\Gamma}\right)^2\right]
\end{equation}
was used again. Here, $\omega_f$ denotes the excitation energy of the final state $f$ and $\Gamma$ the half width at half maximum (HWHM). In accordance with the experiment, a HWHM ($=\frac{1}{2}\times$FWHM) of $0.1$~eV was used throughout.

\begin{acknowledgement}

Funding by the German Research Foundation (DFG) through the collaborative research
center CRC 1249 "N-Heteropolycycles as Functional Materials" (Project Number 281029004-
SFB 1249, projects A01, B01, and B06) is gratefully acknowledged.

\end{acknowledgement}

\begin{suppinfo}

Concentration dependent 2PA measurements, z-scan traces of BT, reversed saturable absorption data, calculated excited-state properties and attachment/detachment densities

\end{suppinfo}

\bibliography{2PA_lit}{}

\providecommand{\latin}[1]{#1}
\makeatletter
\providecommand{\doi}
  {\begingroup\let\do\@makeother\dospecials
  \catcode`\{=1 \catcode`\}=2 \doi@aux}
\providecommand{\doi@aux}[1]{\endgroup\texttt{#1}}
\makeatother
\providecommand*\mcitethebibliography{\thebibliography}
\csname @ifundefined\endcsname{endmcitethebibliography}
  {\let\endmcitethebibliography\endthebibliography}{}
\begin{mcitethebibliography}{30}
\providecommand*\natexlab[1]{#1}
\providecommand*\mciteSetBstSublistMode[1]{}
\providecommand*\mciteSetBstMaxWidthForm[2]{}
\providecommand*\mciteBstWouldAddEndPuncttrue
  {\def\EndOfBibitem{\unskip.}}
\providecommand*\mciteBstWouldAddEndPunctfalse
  {\let\EndOfBibitem\relax}
\providecommand*\mciteSetBstMidEndSepPunct[3]{}
\providecommand*\mciteSetBstSublistLabelBeginEnd[3]{}
\providecommand*\EndOfBibitem{}
\mciteSetBstSublistMode{f}
\mciteSetBstMaxWidthForm{subitem}{(\alph{mcitesubitemcount})}
\mciteSetBstSublistLabelBeginEnd
  {\mcitemaxwidthsubitemform\space}
  {\relax}
  {\relax}

\bibitem[Leray \latin{et~al.}(2008)Leray, Lillis, and Mertz]{bioimaging}
Leray,~A.; Lillis,~K.; Mertz,~J. Enhanced background rejection in thick tissue
  with differential-aberration two-photon microscopy. \emph{Biophysical
  journal} \textbf{2008}, \emph{94}, 1449--1458\relax
\mciteBstWouldAddEndPuncttrue
\mciteSetBstMidEndSepPunct{\mcitedefaultmidpunct}
{\mcitedefaultendpunct}{\mcitedefaultseppunct}\relax
\EndOfBibitem
\bibitem[Kim and Cho(2015)Kim, and Cho]{bioimg1}
Kim,~H.~M.; Cho,~B.~R. Small-molecule two-photon probes for bioimaging
  applications. \emph{Chemical reviews} \textbf{2015}, \emph{115},
  5014--5055\relax
\mciteBstWouldAddEndPuncttrue
\mciteSetBstMidEndSepPunct{\mcitedefaultmidpunct}
{\mcitedefaultendpunct}{\mcitedefaultseppunct}\relax
\EndOfBibitem
\bibitem[Li \latin{et~al.}(2015)Li, Cheng, Huang, Li, and Zhu]{bioimg2}
Li,~J.; Cheng,~F.; Huang,~H.; Li,~L.; Zhu,~J.-J. Nanomaterial-based activatable
  imaging probes: from design to biological applications. \emph{Chemical
  Society reviews} \textbf{2015}, \emph{44}, 7855--7880\relax
\mciteBstWouldAddEndPuncttrue
\mciteSetBstMidEndSepPunct{\mcitedefaultmidpunct}
{\mcitedefaultendpunct}{\mcitedefaultseppunct}\relax
\EndOfBibitem
\bibitem[Park \latin{et~al.}(2015)Park, Lee, Suh, and Hyeon]{bioimg3}
Park,~Y.~I.; Lee,~K.~T.; Suh,~Y.~D.; Hyeon,~T. Upconverting nanoparticles: a
  versatile platform for wide-field two-photon microscopy and multi-modal in
  vivo imaging. \emph{Chemical Society reviews} \textbf{2015}, \emph{44},
  1302--1317\relax
\mciteBstWouldAddEndPuncttrue
\mciteSetBstMidEndSepPunct{\mcitedefaultmidpunct}
{\mcitedefaultendpunct}{\mcitedefaultseppunct}\relax
\EndOfBibitem
\bibitem[Eichelmann \latin{et~al.}(2023)Eichelmann, Monti, Hsu, Kr{\"o}ger,
  Ballmann, Blasco, and Gade]{nanofab}
Eichelmann,~R.; Monti,~J.; Hsu,~L.-Y.; Kr{\"o}ger,~F.; Ballmann,~J.;
  Blasco,~E.; Gade,~L.~H. Two-photon microprinting of 3D emissive structures
  using tetraazaperylene-derived fluorophores. \emph{Molecular Systems Design
  {\&} Engineering} \textbf{2023}, \relax
\mciteBstWouldAddEndPunctfalse
\mciteSetBstMidEndSepPunct{\mcitedefaultmidpunct}
{}{\mcitedefaultseppunct}\relax
\EndOfBibitem
\bibitem[McKenzie \latin{et~al.}(2017)McKenzie, Sazanovich, Baggaley, Bonneau,
  Guerchais, Williams, Weinstein, and Bryant]{phototherapy}
McKenzie,~L.~K.; Sazanovich,~I.~V.; Baggaley,~E.; Bonneau,~M.; Guerchais,~V.;
  Williams,~J. A.~G.; Weinstein,~J.~A.; Bryant,~H.~E. Metal Complexes for
  Two-Photon Photodynamic Therapy: A Cyclometallated Iridium Complex Induces
  Two-Photon Photosensitization of Cancer Cells under Near-IR Light.
  \emph{Chemistry – A European Journal} \textbf{2017}, \emph{23},
  234--238\relax
\mciteBstWouldAddEndPuncttrue
\mciteSetBstMidEndSepPunct{\mcitedefaultmidpunct}
{\mcitedefaultendpunct}{\mcitedefaultseppunct}\relax
\EndOfBibitem
\bibitem[Corredor \latin{et~al.}(2006)Corredor, Huang, and
  Belfield]{datastorage}
Corredor,~C.~C.; Huang,~Z.-L.; Belfield,~K.~D. Two--Photon 3D Optical Data
  Storage via Fluorescence Modulation of an Efficient Fluorene Dye by a
  Photochromic Diarylethene. \emph{Advanced Materials} \textbf{2006},
  \emph{18}, 2910--2914\relax
\mciteBstWouldAddEndPuncttrue
\mciteSetBstMidEndSepPunct{\mcitedefaultmidpunct}
{\mcitedefaultendpunct}{\mcitedefaultseppunct}\relax
\EndOfBibitem
\bibitem[Intorp \latin{et~al.}(2020)Intorp, Hodecker, Müller, Tverskoy,
  Rosenkranz, Dmitrieva, Popov, Rominger, Freudenberg, Dreuw, and
  Bunz]{bunzdiradical}
Intorp,~S.~N.; Hodecker,~M.; Müller,~M.; Tverskoy,~O.; Rosenkranz,~M.;
  Dmitrieva,~E.; Popov,~A.~A.; Rominger,~F.; Freudenberg,~J.; Dreuw,~A.;
  Bunz,~U. H.~F. Quinoidal Azaacenes: 99 \% Diradical Character.
  \emph{Angewandte Chemie International Edition} \textbf{2020}, \emph{59},
  12396--12401\relax
\mciteBstWouldAddEndPuncttrue
\mciteSetBstMidEndSepPunct{\mcitedefaultmidpunct}
{\mcitedefaultendpunct}{\mcitedefaultseppunct}\relax
\EndOfBibitem
\bibitem[Fuchs \latin{et~al.}(2024)Fuchs, Oberhof, Sauter, Pollien, Brödner,
  Rominger, Freudenberg, Dreuw, Tegeder, and Bunz]{Azaacene}
Fuchs,~K.; Oberhof,~N.; Sauter,~G.; Pollien,~A.; Brödner,~K.; Rominger,~F.;
  Freudenberg,~J.; Dreuw,~A.; Tegeder,~P.; Bunz,~U. H.~F. Azaacene Diradicals
  Based on Non-Kekulé Meta- Quinodimethane. \emph{ChemRxiv} \textbf{2024},
  \emph{DOI:10.26434/chemrxiv-2024-77p4z}\relax
\mciteBstWouldAddEndPuncttrue
\mciteSetBstMidEndSepPunct{\mcitedefaultmidpunct}
{\mcitedefaultendpunct}{\mcitedefaultseppunct}\relax
\EndOfBibitem
\bibitem[Takahashi \latin{et~al.}(1996)Takahashi, Gunji, Yanagi, and
  Miki]{bithio}
Takahashi,~K.; Gunji,~A.; Yanagi,~K.; Miki,~M. Synthesis of Novel
  Heteroquaterphenoquinones and Their Electrochemical, Structural, and
  Spectroscopic Characterization. \emph{The Journal of Organic Chemistry}
  \textbf{1996}, \emph{61}, 4784--4792, PMID: 11667412\relax
\mciteBstWouldAddEndPuncttrue
\mciteSetBstMidEndSepPunct{\mcitedefaultmidpunct}
{\mcitedefaultendpunct}{\mcitedefaultseppunct}\relax
\EndOfBibitem
\bibitem[Chung \latin{et~al.}(2005)Chung, Rumi, Alain, Barlow, Perry, and
  Marder]{record2pa}
Chung,~S.-J.; Rumi,~M.; Alain,~V.; Barlow,~S.; Perry,~J.~W.; Marder,~S.~R.
  Strong, low-energy two-photon absorption in extended amine-terminated
  cyano-substituted phenylenevinylene oligomers. \emph{Journal of the American
  Chemical Society} \textbf{2005}, \emph{127}, 10844--10845\relax
\mciteBstWouldAddEndPuncttrue
\mciteSetBstMidEndSepPunct{\mcitedefaultmidpunct}
{\mcitedefaultendpunct}{\mcitedefaultseppunct}\relax
\EndOfBibitem
\bibitem[He \latin{et~al.}(2008)He, Tan, Zheng, and Prasad]{overview2pa}
He,~G.~S.; Tan,~L.-S.; Zheng,~Q.; Prasad,~P.~N. Multiphoton absorbing
  materials: molecular designs, characterizations, and applications.
  \emph{Chemical reviews} \textbf{2008}, \emph{108}, 1245--1330\relax
\mciteBstWouldAddEndPuncttrue
\mciteSetBstMidEndSepPunct{\mcitedefaultmidpunct}
{\mcitedefaultendpunct}{\mcitedefaultseppunct}\relax
\EndOfBibitem
\bibitem[Wang \latin{et~al.}(2001)Wang, Macak, Luo, and Ågren]{Wang_2001}
Wang,~C.-K.; Macak,~P.; Luo,~Y.; Ågren,~H. {Effects of $\pi$ centers and
  symmetry on two-photon absorption cross sections of organic chromophores}.
  \emph{J. Chem. Phys.} \textbf{2001}, \emph{114}, 9813--9820\relax
\mciteBstWouldAddEndPuncttrue
\mciteSetBstMidEndSepPunct{\mcitedefaultmidpunct}
{\mcitedefaultendpunct}{\mcitedefaultseppunct}\relax
\EndOfBibitem
\bibitem[Namboodiri \latin{et~al.}(2016)Namboodiri, Bongu, Bisht, Mukkamala,
  Chandra, Aidhen, Kelly, and Costello]{ICT}
Namboodiri,~C.; Bongu,~S.~R.; Bisht,~P.~B.; Mukkamala,~R.; Chandra,~B.;
  Aidhen,~I.~S.; Kelly,~T.~J.; Costello,~J.~T. Enhanced two photon absorption
  cross section and optical nonlinearity of a quasi-octupolar molecule.
  \emph{Journal of photochemistry and photobiology. A, Chemistry}
  \textbf{2016}, \emph{314}, 60--65\relax
\mciteBstWouldAddEndPuncttrue
\mciteSetBstMidEndSepPunct{\mcitedefaultmidpunct}
{\mcitedefaultendpunct}{\mcitedefaultseppunct}\relax
\EndOfBibitem
\bibitem[Vivas \latin{et~al.}(2014)Vivas, Silva, Malinge, Boujtita,
  Zale{\'s}ny, Bartkowiak, {\AA}gren, Canuto, de~Boni, Ishow, and
  Mendonca]{ICTPI-Bi}
Vivas,~M.~G.; Silva,~D.~L.; Malinge,~J.; Boujtita,~M.; Zale{\'s}ny,~R.;
  Bartkowiak,~W.; {\AA}gren,~H.; Canuto,~S.; de~Boni,~L.; Ishow,~E.;
  Mendonca,~C.~R. Molecular structure-optical property relationships for a
  series of non-centrosymmetric two-photon absorbing push-pull triarylamine
  molecules. \emph{Scientific reports} \textbf{2014}, \emph{4}, 4447\relax
\mciteBstWouldAddEndPuncttrue
\mciteSetBstMidEndSepPunct{\mcitedefaultmidpunct}
{\mcitedefaultendpunct}{\mcitedefaultseppunct}\relax
\EndOfBibitem
\bibitem[Richy \latin{et~al.}(2023)Richy, Gam, Messaoudi, Triadon, Mongin,
  Blanchard-Desce, Latouche, Humphrey, Boucekkine, Halet, and Paul]{ICT-Bi}
Richy,~N.; Gam,~S.; Messaoudi,~S.; Triadon,~A.; Mongin,~O.;
  Blanchard-Desce,~M.; Latouche,~C.; Humphrey,~M.~G.; Boucekkine,~A.;
  Halet,~J.-F.; Paul,~F. Linear and Nonlinear Optical Properties of Quadrupolar
  Bithiophenes and Cyclopentadithiophenes as Fluorescent Oxygen
  Photosensitizers. \emph{Photochem} \textbf{2023}, \emph{3}, 127--154\relax
\mciteBstWouldAddEndPuncttrue
\mciteSetBstMidEndSepPunct{\mcitedefaultmidpunct}
{\mcitedefaultendpunct}{\mcitedefaultseppunct}\relax
\EndOfBibitem
\bibitem[Beerepoot \latin{et~al.}(2018)Beerepoot, Alam, Bednarska, Bartkowiak,
  Ruud, and Zaleśny]{Beerepoot_2018}
Beerepoot,~M. T.~P.; Alam,~M.~M.; Bednarska,~J.; Bartkowiak,~W.; Ruud,~K.;
  Zaleśny,~R. Benchmarking the Performance of Exchange-Correlation Functionals
  for Predicting Two-Photon Absorption Strengths. \emph{J. Chem. Theory
  Comput.} \textbf{2018}, \emph{14}, 3677--3685, PMID: 29852067\relax
\mciteBstWouldAddEndPuncttrue
\mciteSetBstMidEndSepPunct{\mcitedefaultmidpunct}
{\mcitedefaultendpunct}{\mcitedefaultseppunct}\relax
\EndOfBibitem
\bibitem[Cronstrand \latin{et~al.}(2002)Cronstrand, Luo, and
  Ågren]{Cronstrand_2002}
Cronstrand,~P.; Luo,~Y.; Ågren,~H. Generalized few-state models for two-photon
  absorption of conjugated molecules. \emph{Chem. Phys. Lett.} \textbf{2002},
  \emph{352}, 262--269\relax
\mciteBstWouldAddEndPuncttrue
\mciteSetBstMidEndSepPunct{\mcitedefaultmidpunct}
{\mcitedefaultendpunct}{\mcitedefaultseppunct}\relax
\EndOfBibitem
\bibitem[Lin \latin{et~al.}(2011)Lin, Luo, Ruud, Zhao, Santoro, and
  Rizzo]{Lin_2011}
Lin,~N.; Luo,~Y.; Ruud,~K.; Zhao,~X.; Santoro,~F.; Rizzo,~A. Differences in
  Two-Photon and One-Photon Absorption Profiles Induced by Vibronic Coupling:
  The Case of Dioxaborine Heterocyclic Dye. \emph{ChemPhysChem} \textbf{2011},
  \emph{12}, 3392--3403\relax
\mciteBstWouldAddEndPuncttrue
\mciteSetBstMidEndSepPunct{\mcitedefaultmidpunct}
{\mcitedefaultendpunct}{\mcitedefaultseppunct}\relax
\EndOfBibitem
\bibitem[Beerepoot \latin{et~al.}(2015)Beerepoot, Friese, List, Kongsted, and
  Ruud]{Beerepoot_2015}
Beerepoot,~M. T.~P.; Friese,~D.~H.; List,~N.~H.; Kongsted,~J.; Ruud,~K.
  Benchmarking two-photon absorption cross sections: performance of CC2 and
  CAM-B3LYP. \emph{Phys. Chem. Chem. Phys.} \textbf{2015}, \emph{17},
  19306--19314\relax
\mciteBstWouldAddEndPuncttrue
\mciteSetBstMidEndSepPunct{\mcitedefaultmidpunct}
{\mcitedefaultendpunct}{\mcitedefaultseppunct}\relax
\EndOfBibitem
\bibitem[Sutherland(1996)]{Sutherland.1996}
Sutherland,~R.~L. \emph{Handbook of Nonlinear Optics}; Optical Engineering.
  Series Editor; {Marcel Dekker}: New York, 1996; Vol.~52\relax
\mciteBstWouldAddEndPuncttrue
\mciteSetBstMidEndSepPunct{\mcitedefaultmidpunct}
{\mcitedefaultendpunct}{\mcitedefaultseppunct}\relax
\EndOfBibitem
\bibitem[Sheik-Bahae \latin{et~al.}(1990)Sheik-Bahae, Said, Wei, Hagan, and
  {van Stryland}]{SheikBahae.1990}
Sheik-Bahae,~M.; Said,~A.~A.; Wei,~T.-H.; Hagan,~D.~J.; {van Stryland},~E.~W.
  Sensitive measurement of optical nonlinearities using a single beam.
  \emph{IEEE Journal of Quantum Electronics} \textbf{1990}, \emph{26},
  760--769\relax
\mciteBstWouldAddEndPuncttrue
\mciteSetBstMidEndSepPunct{\mcitedefaultmidpunct}
{\mcitedefaultendpunct}{\mcitedefaultseppunct}\relax
\EndOfBibitem
\bibitem[Epifanovsky \latin{et~al.}(2021)Epifanovsky, Gilbert, Feng, Lee, Mao,
  Mardirossian, Pokhilko, White, Coons, Dempwolff, Gan, Hait, Horn, Jacobson,
  Kaliman, Kussmann, Lange, Lao, Levine, Liu, McKenzie, Morrison, Nanda,
  Plasser, Rehn, Vidal, You, Zhu, Alam, Albrecht, Aldossary, Alguire, Andersen,
  Athavale, Barton, Begam, Behn, Bellonzi, Bernard, Berquist, Burton, Carreras,
  Carter-Fenk, Chakraborty, Chien, Closser, Cofer-Shabica, Dasgupta,
  de~Wergifosse, Deng, Diedenhofen, Do, Ehlert, Fang, Fatehi, Feng, Friedhoff,
  Gayvert, Ge, Gidofalvi, Goldey, Gomes, González-Espinoza, Gulania, Gunina,
  Hanson-Heine, Harbach, Hauser, Herbst, Hernández~Vera, Hodecker, Holden,
  Houck, Huang, Hui, Huynh, Ivanov, Jász, Ji, Jiang, Kaduk, Kähler,
  Khistyaev, Kim, Kis, Klunzinger, Koczor-Benda, Koh, Kosenkov, Koulias,
  Kowalczyk, Krauter, Kue, Kunitsa, Kus, Ladjánszki, Landau, Lawler,
  Lefrancois, Lehtola, Li, Li, Liang, Liebenthal, Lin, Lin, Liu, Liu,
  Loipersberger, Luenser, Manjanath, Manohar, Mansoor, Manzer, Mao, Marenich,
  Markovich, Mason, Maurer, McLaughlin, Menger, Mewes, Mewes, Morgante,
  Mullinax, Oosterbaan, Paran, Paul, Paul, Pavošević, Pei, Prager, Proynov,
  Rák, Ramos-Cordoba, Rana, Rask, Rettig, Richard, Rob, Rossomme, Scheele,
  Scheurer, Schneider, Sergueev, Sharada, Skomorowski, Small, Stein, Su,
  Sundstrom, Tao, Thirman, Tornai, Tsuchimochi, Tubman, Veccham, Vydrov,
  Wenzel, Witte, Yamada, Yao, Yeganeh, Yost, Zech, Zhang, Zhang, Zhang, Zuev,
  Aspuru-Guzik, Bell, Besley, Bravaya, Brooks, Casanova, Chai, Coriani, Cramer,
  Cserey, DePrince, DiStasio, Dreuw, Dunietz, Furlani, Goddard,
  Hammes-Schiffer, Head-Gordon, Hehre, Hsu, Jagau, Jung, Klamt, Kong,
  Lambrecht, Liang, Mayhall, McCurdy, Neaton, Ochsenfeld, Parkhill, Peverati,
  Rassolov, Shao, Slipchenko, Stauch, Steele, Subotnik, Thom, Tkatchenko,
  Truhlar, Van~Voorhis, Wesolowski, Whaley, Woodcock, Zimmerman, Faraji, Gill,
  Head-Gordon, Herbert, and Krylov]{Epifanovsky_2021}
Epifanovsky,~E. \latin{et~al.}  {Software for the frontiers of quantum
  chemistry: An overview of developments in the Q-Chem 5 package}. \emph{J.
  Chem. Phys.} \textbf{2021}, \emph{155}, 084801\relax
\mciteBstWouldAddEndPuncttrue
\mciteSetBstMidEndSepPunct{\mcitedefaultmidpunct}
{\mcitedefaultendpunct}{\mcitedefaultseppunct}\relax
\EndOfBibitem
\bibitem[Liu and Liang(2013)Liu, and Liang]{Liu_2013}
Liu,~J.; Liang,~W. {Analytical second derivatives of excited-state energy
  within the time-dependent density functional theory coupled with a
  conductor-like polarizable continuum model}. \emph{J. Chem. Phys.}
  \textbf{2013}, \emph{138}, 024101\relax
\mciteBstWouldAddEndPuncttrue
\mciteSetBstMidEndSepPunct{\mcitedefaultmidpunct}
{\mcitedefaultendpunct}{\mcitedefaultseppunct}\relax
\EndOfBibitem
\bibitem[Yanai \latin{et~al.}(2004)Yanai, Tew, and Handy]{Yanai_2004}
Yanai,~T.; Tew,~D.~P.; Handy,~N.~C. A new hybrid exchange–correlation
  functional using the Coulomb-attenuating method (CAM-B3LYP). \emph{Chem.
  Phys. Lett.} \textbf{2004}, \emph{393}, 51--57\relax
\mciteBstWouldAddEndPuncttrue
\mciteSetBstMidEndSepPunct{\mcitedefaultmidpunct}
{\mcitedefaultendpunct}{\mcitedefaultseppunct}\relax
\EndOfBibitem
\bibitem[Grimme \latin{et~al.}(2010)Grimme, Antony, Ehrlich, and
  Krieg]{Grimme_2010}
Grimme,~S.; Antony,~J.; Ehrlich,~S.; Krieg,~H. {A consistent and accurate ab
  initio parametrization of density functional dispersion correction (DFT-D)
  for the 94 elements H-Pu}. \emph{J. Chem. Phys.} \textbf{2010}, \emph{132},
  154104\relax
\mciteBstWouldAddEndPuncttrue
\mciteSetBstMidEndSepPunct{\mcitedefaultmidpunct}
{\mcitedefaultendpunct}{\mcitedefaultseppunct}\relax
\EndOfBibitem
\bibitem[Grimme \latin{et~al.}(2011)Grimme, Ehrlich, and Goerigk]{Grimme_2011}
Grimme,~S.; Ehrlich,~S.; Goerigk,~L. Effect of the damping function in
  dispersion corrected density functional theory. \emph{J. Comput. Chem.}
  \textbf{2011}, \emph{32}, 1456--1465\relax
\mciteBstWouldAddEndPuncttrue
\mciteSetBstMidEndSepPunct{\mcitedefaultmidpunct}
{\mcitedefaultendpunct}{\mcitedefaultseppunct}\relax
\EndOfBibitem
\bibitem[Woon and Dunning(1993)Woon, and Dunning]{Woon_1993}
Woon,~D.~E.; Dunning,~J.,~Thom~H. {Gaussian basis sets for use in correlated
  molecular calculations. III. The atoms aluminum through argon}. \emph{J.
  Chem. Phys.} \textbf{1993}, \emph{98}, 1358--1371\relax
\mciteBstWouldAddEndPuncttrue
\mciteSetBstMidEndSepPunct{\mcitedefaultmidpunct}
{\mcitedefaultendpunct}{\mcitedefaultseppunct}\relax
\EndOfBibitem
\bibitem[Hirata and Head-Gordon(1999)Hirata, and Head-Gordon]{Hirata_1999}
Hirata,~S.; Head-Gordon,~M. Time-dependent density functional theory within the
  Tamm–Dancoff approximation. \emph{Chem. Phys. Lett.} \textbf{1999},
  \emph{314}, 291--299\relax
\mciteBstWouldAddEndPuncttrue
\mciteSetBstMidEndSepPunct{\mcitedefaultmidpunct}
{\mcitedefaultendpunct}{\mcitedefaultseppunct}\relax
\EndOfBibitem
\end{mcitethebibliography}

\clearpage
 \pagenumbering{arabic}
\section{Supplementary Material}
\beginsupplement

\subsection{Concentration Dependent 2PA Measurements}

\begin{figure}
\begin{center}
\includegraphics[width=16cm]{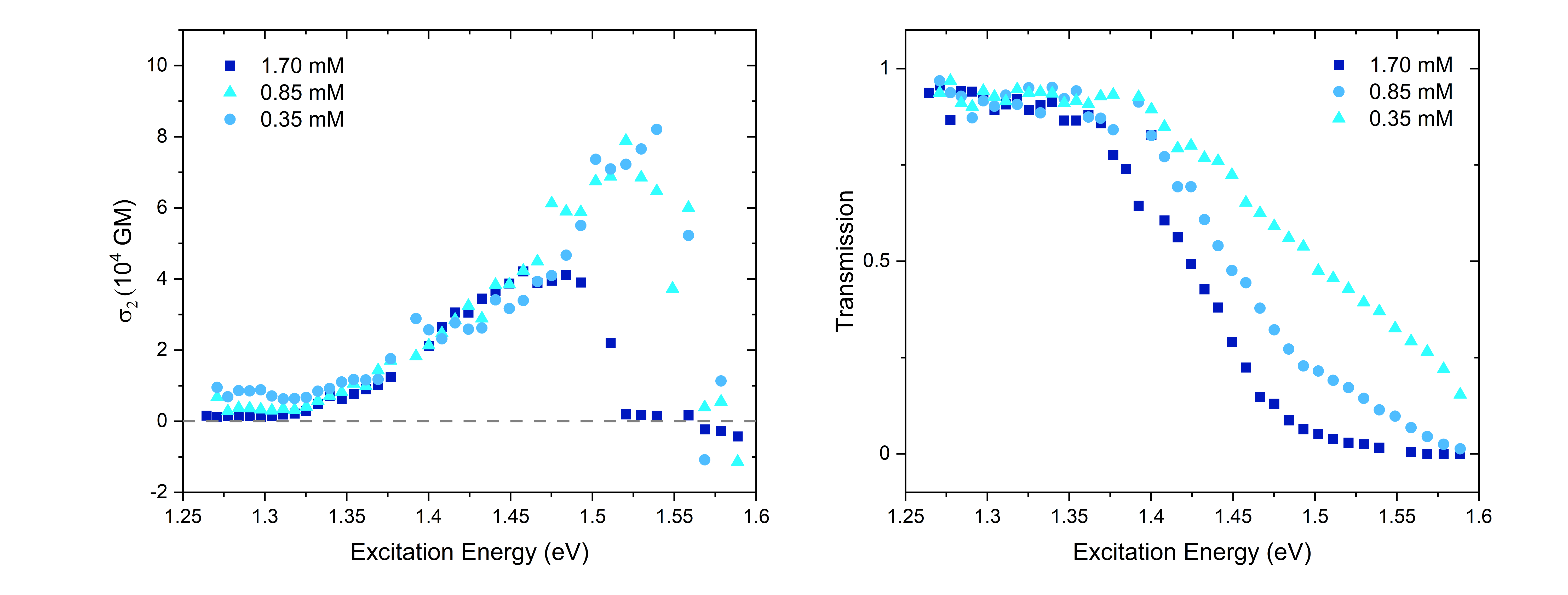}
\caption{In a) one can see the 2PA spectra measured for different concentrations of BT solutions. In b) the linear transmission of the beam for each solution is plotted. }
\label{fig:S-BT}
\end{center}
\end{figure} 

Concentration-dependent measurements provide insights into the interplay between linear and nonlinear optical absorption. As depicted in Figure \ref{fig:S-BT}b, the onset of linear absorption is observed at approximately 1.4 eV. According to the Lambert-Beer law, solutions with higher concentrations exhibit absorption at lower photon energies compared to those with lower concentrations. Consequently, the nonlinear effect of saturable absorption (SA) is initiated  as competing effect at an earlier stage in more concentrated solutions. Conversely, in solutions with lower concentrations, two-photon absorption can still be measured at higher energies. However, in lower concentrated solutions the effect of saturable absorption is more pronounced due to the reduced number of molecules, which leads to a more rapid onset of saturation. This dynamic results in a steeper decline in nonlinear absorption. The observed trend culminates in the predominance of reverse saturable absorption at higher photon energies, as shown in Figure \ref{fig:S-BT} a). This behavior underscores the complex relationship between molecule concentration, linear absorption, and the competing nonlinear effects within the BT solutions.
\\

\begin{figure}
\begin{center}
\includegraphics[width=16cm]{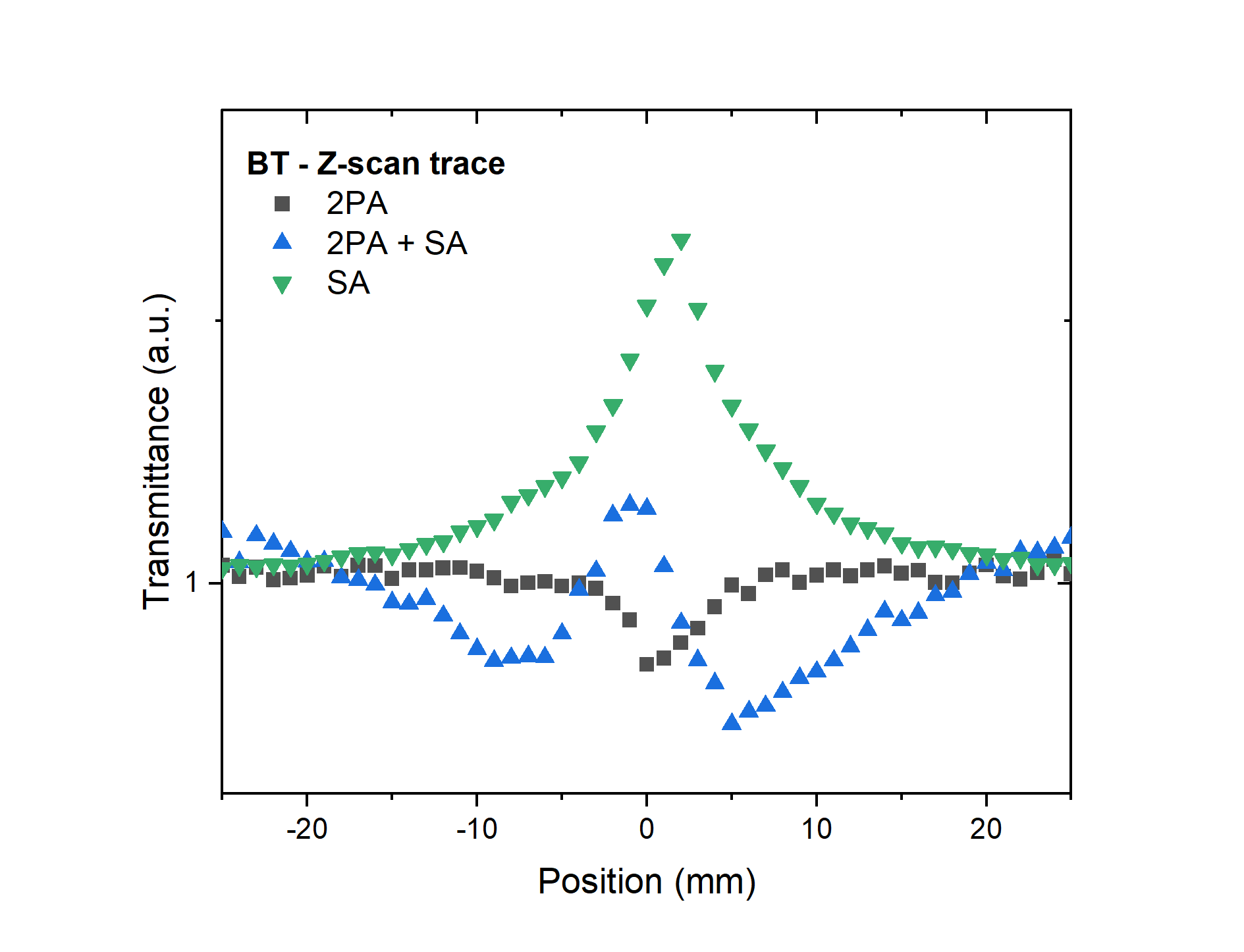}
\caption{Exemplary z-scan traces of BT. They show just 2PA, 2PA while already SA present and mainly SA.}
\label{fig:S-BT}
\end{center}
\end{figure} 

\subsection{Reversed Saturable Absorption} \label{sec:SI-RSA}
\begin{figure}
\begin{center}
\includegraphics[width=16cm]{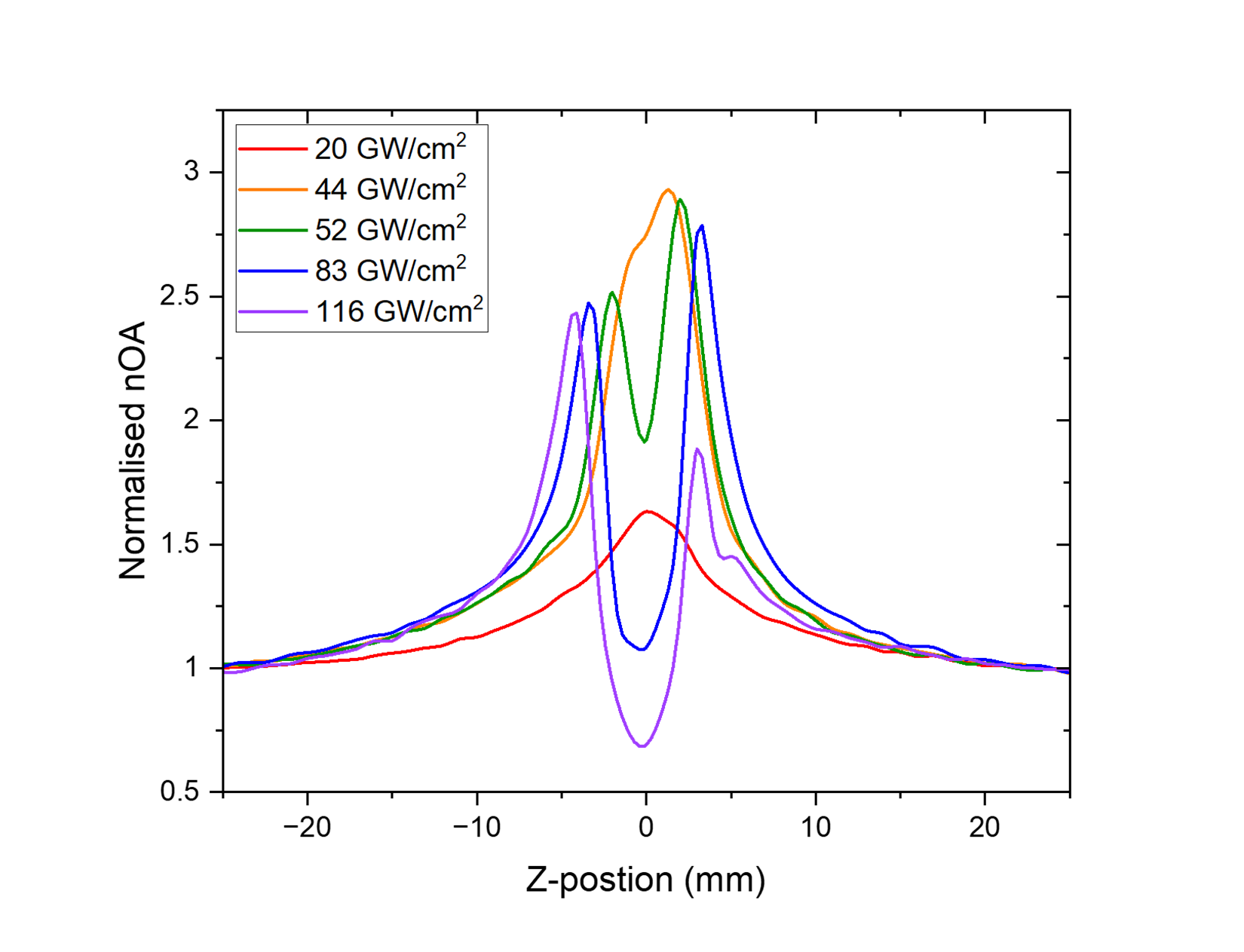}
\caption{Z-san traces of BT with different input peak intensities at a fixed excitation energy (wavelength).}
\label{fig:S-BT-rsa}
\end{center}
\end{figure}

The dependence of NLO response of the bithiophene on the incident intensity in the high energy regime has been investigated by adopting a wide range of input intensities and can be seen in Fig. \ref{fig:S-BT-rsa}. It can be clearly seen that there are overlaying effects, where in the low intensity regime SA effects dominates. When the input peak intensity is increased, a valley in the the normalized transmittance.emerges at the focal point, indicating typical response for reversed saturable absorption (RSA). In this case the RSA is attributed to the large two photon absorption at this excitation energy. It is apparent that SA and nonlinear absorption coexist, with the shape of the trace as a result of competing nonlinear effects.%\cite{SaturableAbs}

\subsection{Excited-State Properties and Attachment/Detachment Densities}
\subsubsection{BT}

\begin{figure}[H]
\begin{center}
\includegraphics[width=0.9\textwidth]{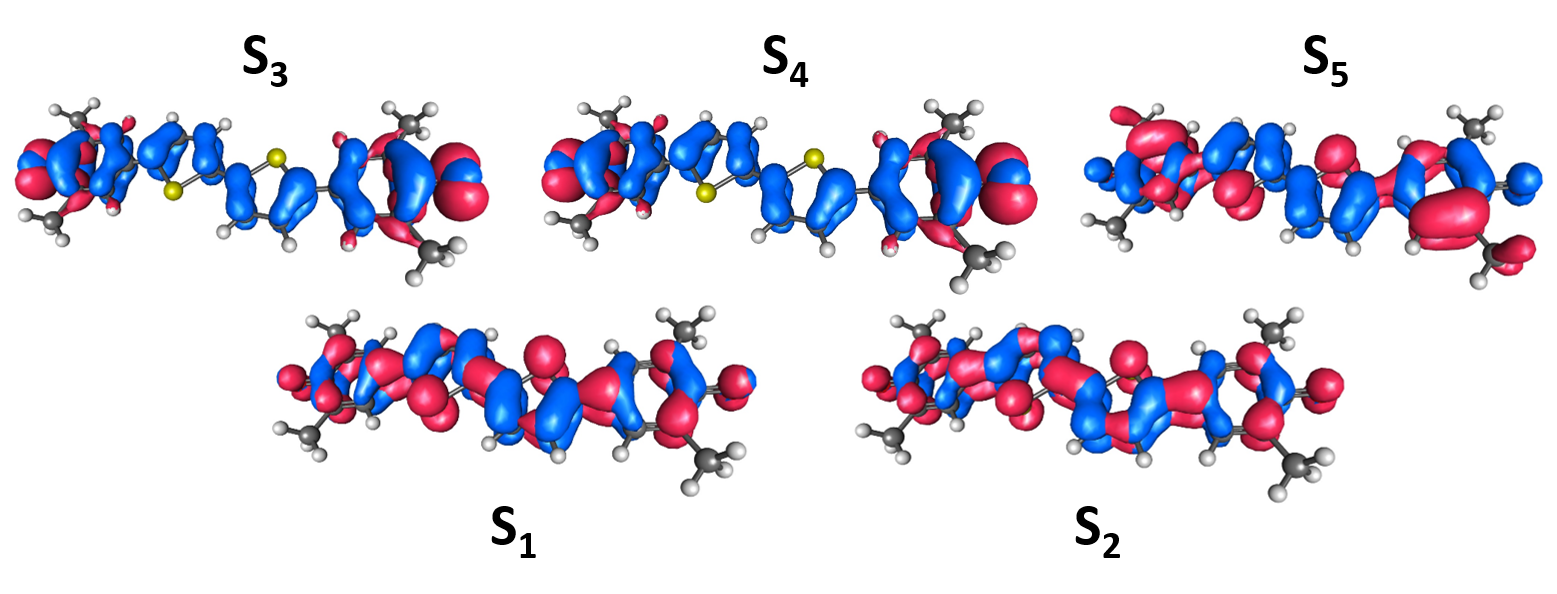}
\caption{Attachment/detachment densities in blue/red of the first 5 excited singlet states of BT calculated at the CAM-B3LYP-D3(BJ)/aug-cc-pVDZ level of theory.}
\label{fig:ad_BT_5}
\end{center}
\end{figure} 

\subsubsection{DA-BT}
\begin{table}[H]
\caption{Excitation energies, oscillator strengths, 2PA cross sections, transition moments from the ground state, difference dipole moments, and state-to-state transition moments for the first 8 excited singlet states of DA-BT calculated at the CAM-B3LYP-D3(BJ)/aug-cc-pVDZ level of theory.}
\label{tbl:sos_dafbt}
\begin{tabular}{ccccccc}
\toprule
%\multicolumn{7}{c}{\textbf{DA-f-BT}}\\
%\midrule
$f$ & $\omega_f$ (ev) & $f_L$ (a.u.) & $\sigma_2$ (GM) & $\vert\braket{0|\vec{\mu}|f}\rvert$ (a.u.) & $\lvert\braket{f|\vec{\overline{\mu}}|f}\rvert$ (a.u.) & $\lvert\braket{1|\vec{\mu}|f}\rvert$ (a.u.)\\
\midrule
1 & 1.86 & 3.495 & 46 & 8.758 & 1.124 & - \\
2 & 2.77 & 0.003 & 35857 & 0.193 & 2.390 & 6.466 \\
3 & 2.95 & 0.000 & 0 & 0.015 & 3.603 & 0.004 \\
4 & 2.95 & 0.000 & 0 & 0.016 & 3.601 & 0.007 \\
5 & 3.00 & 0.334 & 199 & 2.131 & 1.796 & 0.481 \\
6 & 3.14 & 0.208 & 36783 & 1.647 & 2.987 & 3.557 \\
7 & 3.33 & 0.067 & 7472 & 0.910 & 2.265 & 0.956 \\
8 & 3.35 & 0.007 & 2942 & 0.284 & 1.057 & 0.957 \\
\bottomrule
\end{tabular}
\end{table}

\subsubsection{DA(n)-BT}
\begin{table}[H]
\caption{Excitation energies, oscillator strengths, 2PA cross sections, transition moments from the ground state, difference dipole moments, and state-to-state transition moments for the first 8 excited singlet states of DA(n)-BT calculated at the CAM-B3LYP-D3(BJ)/aug-cc-pVDZ level of theory.}
\label{tbl:sos_danfbt}
\begin{tabular}{ccccccc}
\toprule
\multicolumn{7}{c}{\textbf{DAA-BT}}\\
\midrule
$f$ & $\omega_f$ (ev) & $f_L$ (a.u.) & $\sigma_2$ (GM) & $\vert\braket{0|\vec{\mu}|f}\rvert$ (a.u.) & $\lvert\braket{f|\vec{\overline{\mu}}|f}\rvert$ (a.u.) & $\lvert\braket{1|\vec{\mu}|f}\rvert$ (a.u.)\\
\midrule
1 & 1.88 & 3.249 & 39 & 8.392 & 1.084 & - \\
2 & 2.26 & 0.067 & 2011 & 1.099 & 7.526 & 3.161 \\
3 & 2.90 & 0.297 & 28751 & 2.043 & 3.831 & 5.205 \\
4 & 2.97 & 0.000 & 0 & 0.004 & 4.533 & 0.006 \\
5 & 2.97 & 0.000 & 0 & 0.004 & 4.533 & 0.009 \\
6 & 3.00 & 0.001 & 0 & 0.134 & 2.938 & 0.007 \\
7 & 3.08 & 0.012 & 204 & 0.401 & 2.176 & 0.542 \\
8 & 3.11 & 0.006 & 44351 & 0.284 & 3.870 & 4.575 \\
\midrule
\multicolumn{7}{c}{\textbf{DAT-BT}}\\
\midrule
$f$ & $\omega_f$ (ev) & $f_L$ (a.u.) & $\sigma_2$ (GM) & $\vert\braket{0|\vec{\mu}|f}\rvert$ (a.u.) & $\lvert\braket{f|\vec{\overline{\mu}}|f}\rvert$ (a.u.) & $\lvert\braket{1|\vec{\mu}|f}\rvert$ (a.u.)\\
\midrule
1 & 1.90 & 3.179 & 43 & 8.265 & 1.097 & - \\
2 & 2.12 & 0.084 & 1304 & 1.269 & 7.854 & 3.043 \\
3 & 2.44 & 0.307 & 971 & 2.265 & 9.881 & 1.677 \\
4 & 2.53 & 0.302 & 1 & 2.207 & 3.296 & 0.304 \\
5 & 2.99 & 0.000 & 0 & 0.008 & 4.762 & 0.014 \\
6 & 2.99 & 0.000 & 0 & 0.002 & 4.762 & 0.005 \\
7 & 3.04 & 0.111 & 69993 & 1.219 & 0.734 & 6.823 \\
8 & 3.16 & 0.311 & 371 & 2.003 & 2.885 & 0.566 \\
\midrule
\multicolumn{7}{c}{\textbf{DAP-BT}}\\
\midrule
$f$ & $\omega_f$ (ev) & $f_L$ (a.u.) & $\sigma_2$ (GM) & $\vert\braket{0|\vec{\mu}|f}\rvert$ (a.u.) & $\lvert\braket{f|\vec{\overline{\mu}}|f}\rvert$ (a.u.) & $\lvert\braket{1|\vec{\mu}|f}\rvert$ (a.u.)\\
\midrule
1 & 1.90 & 2.853 & 49 & 7.827 & 0.972 & - \\
2 & 1.99 & 0.070 & 1002 & 1.198 & 8.014 & 2.694 \\
3 & 2.08 & 0.568 & 1 & 3.337 & 3.967 & 0.948 \\
4 & 2.15 & 0.396 & 600 & 2.743 & 10.916 & 0.796 \\
5 & 2.99 & 0.000 & 2 & 0.014 & 6.879 & 0.039 \\
6 & 2.99 & 0.000 & 0 & 0.005 & 6.874 & 0.015 \\
7 & 3.02 & 0.212 & 66066 & 1.690 & 0.866 & 6.867 \\
8 & 3.17 & 0.231 & 370 & 1.724 & 3.291 & 0.534 \\
\bottomrule
\end{tabular}
\end{table}
%\subsection{Analysis of SOS Contributions}
%\begin{figure}[H]
%\begin{center}
%\includegraphics[width=\textwidth]{Figs/sos_BT.eps}
%\caption{.}
%\label{fig:sos_BT}
%\end{center}
%\end{figure} 
%\begin{figure}[H]
%\begin{center}
%\includegraphics[width=\textwidth]{Figs/sos_2Mstar.eps}
%\caption{.}
%\label{fig:sos_DAfBT}
%\end{center}
%\end{figure} 

%\begin{figure}[H]
%\begin{center}
%\includegraphics[width=\textwidth]{Figs/sos_DAAfBT.eps}
%\caption{.}
%\label{fig:sos_DAAfBT}
%\end{center}
%\end{figure} 

%\begin{figure}[H]
%\begin{center}
%\includegraphics[width=\textwidth]{Figs/sos_DATfBT.eps}
%\caption{.}
%\label{fig:sos_DATfBT}
%\end{center}
%\end{figure} 

%\begin{figure}[H]
%\begin{center}
%\includegraphics[width=\textwidth]{Figs/sos_DAPfBT.eps}
%\caption{.}
%\label{fig:sos_DAPfBT}
%\end{center}
%\end{figure} 

\end{document}